\documentclass{llncs}

\usepackage[utf8]{inputenc}
\usepackage[T1]{fontenc}
\usepackage{paralist}
\usepackage{relsize}
\usepackage{xspace}
\usepackage{algorithm}
\usepackage{graphicx}
\usepackage{listings}
\usepackage{caption}

\usepackage{amsmath}
\usepackage{amsfonts}
\usepackage{bbm, dsfont}
\usepackage{mathrsfs}
\usepackage[svgnames,table]{xcolor}
\usepackage{esvect}
\usepackage{xspace}
\usepackage{url}

\newcommand*{\EXTENDEDVERSION}{}

\ifdefined\EXTENDEDVERSION
\usepackage[breaklinks,hidelinks,pdftitle={Static Detection of DoS Vulnerabilities in Programs that use Regular Expressions (Extended Version)},pdfauthor={Valentin Wüstholz, Oswaldo Olivo, Marijn J. H. Heule, Isil Dillig},pdfkeywords={regular expressions, program analysis, security}]{hyperref}\title{Static Detection of DoS Vulnerabilities in Programs that use Regular
  Expressions (Extended Version)}
\else
\usepackage[breaklinks,hidelinks]{hyperref}
\fi
\usepackage{array, multirow} 
\usepackage{rotating, makecell} 
\usepackage[noend]{algpseudocode}
\usepackage{balance}
\usepackage{marvosym}
\usepackage{stmaryrd}
\usepackage{multirow}
\usepackage{fancyvrb}
\usepackage{comment}
\usepackage{wrapfig}
\usepackage{tikz}
\usetikzlibrary{arrows,automata,decorations.pathmorphing}

\newcommand{\den}[1]{\llbracket#1\rrbracket}
\newcommand{\interv}{\mathbb{I}}
\newcommand{\impure}{\mathcal{R}}
\newcommand{\len}[1]{\emph{len}(#1)}
\newcommand{\interval}[2]{\langle #1, #2\rangle}
\newcommand{\evalregex}[2]{{\den{#1}_{#2}}}
\newcommand{\evalint}[2]{{\den{#1}_{#2}}}
\newcommand{\il}{{\sc Strimp}\xspace}

\newcommand{\tool}{{\sc Rexploiter}\xspace}
\newcommand{\regex}{\mathcal{E}}
\newcommand{\automaton}{\mathcal{A}}
\newcommand{\labl}{\ell}
\newcommand{\npath}{\pi}
\newcommand{\labels}{\emph{labels}}
\newcommand{\langu}{\mathcal L}

\newcommand{\evil}{{\text{\Neptune}}}
\newcommand{\attack}{\automaton^\evil}

\newcommand{\ite}[2]{{#1} \otimes {#2}}

\newcommand{\marijn}[1]{\textcolor{magenta}{\textbf{MARIJN:} #1}}
\newcommand{\todo}[1]{\textcolor{red}{#1}}

\newcommand{\irule}[2]%
   {\mkern-2mu\displaystyle\frac{#1}{\vphantom{,}#2}\mkern-2mu}

\tikzstyle{every circle node}=[circle,draw,thick,inner sep=1.5pt,minimum size=7mm]
\tikzset{snake it/.style={-stealth,
decoration={snake,
    amplitude = .8mm,
    segment length = 4mm,
    post length=1mm},decorate}}

\lstdefinestyle{java}{%
  language         = Java,%
  keywordstyle     = \bfseries\color{DarkBlue},%
  commentstyle     = \itshape,%
  basicstyle       = \ttfamily\scriptsize,%
  columns          = [c]fixed,%
  aboveskip        = 2mm,%
  belowskip        = 2mm,%
  keepspaces       = true,%
  mathescape       = true,%
  escapechar       = ¤,%
  tabsize          = 2,%
  numbers          = left,%
  numberstyle      = \tiny\color{Gray},%
  numbersep        = 6pt,%
  stepnumber       = 1,%
  firstnumber      = 1,%
  showstringspaces = false,%
  captionpos       = b,%
  extendedchars    = true,%
  abovecaptionskip = 0mm,%
  belowcaptionskip = 0mm,%
  alsoletter       = {.},%
}

\begin{document}

\mainmatter

\ifdefined\EXTENDEDVERSION
\title{Static Detection of DoS Vulnerabilities in Programs that use Regular
  Expressions (Extended Version)}
\else
\title{Static Detection of DoS Vulnerabilities in Programs that use Regular Expressions}
\fi

\titlerunning{\ }

\author{Valentin W{\"u}stholz \and Oswaldo Olivo \and Marijn J. H. Heule \and Isil Dillig}

\authorrunning{W{\"u}stholz, Olivo, Heule, Dillig}

\institute{
\vspace{-0.1in} The University of Texas at Austin\\
  \email{\{valentin, olivo, marijn, isil\}@cs.utexas.edu}}

\maketitle

\thispagestyle{plain}


\begin{abstract}
\vspace{-0.2in}
In an \emph{algorithmic complexity attack}, a malicious party takes advantage of the worst-case behavior of an algorithm to cause denial-of-service. A prominent algorithmic complexity attack is \emph{regular expression denial-of-service} (\emph{ReDoS}), in which the attacker exploits a vulnerable regular expression by providing a carefully-crafted input string that triggers worst-case behavior of the  matching algorithm. This paper proposes a technique for automatically finding ReDoS vulnerabilities in programs. Specifically, our approach automatically identifies \emph{vulnerable regular expressions} in the program and determines whether an ``evil" input string can be matched against a vulnerable regular expression. We have implemented our proposed approach in a tool called \tool and found 41 exploitable security vulnerabilities in Java web applications. 


\end{abstract}

\section{Introduction}
\label{sec:intro}





Regular expressions provide a versatile mechanism for parsing and validating input data. 
Due to their  flexibility, many developers use regular expressions to validate 
passwords or to extract substrings that match a given pattern. Hence, 
many languages provide  extensive support for regular expression matching.

While there are several  algorithms for determining membership in a regular language, 
a common technique is to construct a non-deterministic finite automaton (NFA) 
and  perform backtracking search over all possible runs of 
this NFA. Although simple and flexible, this  strategy has  super-linear (in fact, exponential) complexity 
and is prone to a class of \emph{algorithmic complexity attacks}~\cite{CrosbyWallach2003}. 
For some regular expressions (e.g., {\tt (a|b)*(a|c)*}), it is  possible to craft input strings that could cause the matching algorithm 
to take quadratic time (or worse) in the size of the input. For some regular expressions (e.g, {\tt (a+)+}), one can even generate input strings that could cause the matching algorithm to take exponential time.
Hence, attackers  exploit the presence of vulnerable regular expressions to launch so-called \emph{regular expression denial-of-service (ReDoS)} attacks.

ReDoS attacks have been shown to severely impact the responsiveness and availability of applications. For example, the .NET framework was shown to be vulnerable to a ReDoS attack  that paralyzed applications using .NET's default validation mechanism~\cite{cve-net}. Furthermore, unlike other
DoS attacks that require thousands of machines to bring down
critical infrastructure, ReDoS attacks can be triggered by a single malicious
user input. 
Consequently, developers are responsible for protecting their code against 
such attacks, either by avoiding the use of vulnerable regular expressions or by
\emph{sanitizing}  user input.


Unfortunately, protecting an application against ReDoS attacks can be non-trivial in practice. Often,
developers do not know which regular expressions are vulnerable or how to rewrite them in a way that avoids  super-linear complexity. In addition, it is difficult to implement a suitable sanitizer
without understanding the class of input strings that trigger worst-case behavior.
Even though some libraries (e.g., the \textsc{.Net} framework) allow developers to set a time limit for regular expression matching, existing solutions do not address the root cause of the problem.
As a result, ReDoS vulnerabilities are still being uncovered in
many important applications.  For instance, according to the National
Vulnerability Database (NVD), 
there are over 150 acknowledged ReDoS vulnerabilities, some of which are caused by
exponential matching complexity (e.g.,~\cite{cve-net,cve-exp3}) and some of
which are characterized by super-linear behavior
(e.g.,~\cite{cve-superlinear1,cve-superlinear2,cve-superlinear3}).

In this paper, we propose a static technique for automatically uncovering  DoS vulnerabilities in 
programs that use regular expressions.  There are two main technical challenges that make this problem 
difficult: First, given a regular expression $\regex$, we need to statically determine 
the worst-case complexity of matching $\regex$ against an arbitrary input string. Second, given an application $A$ that contains 
a vulnerable regular expression $\regex$, we must statically determine whether there can exist an execution of $A$ in which $\regex$ can be matched against an input string that could cause super-linear behavior.

We solve these challenges by developing a two-tier algorithm that combines (a) static analysis of regular expressions with (b) sanitization-aware taint analysis at the source code level. Our technique can identify both \emph{vulnerable} regular expressions that have super-linear  complexity (quadratic or worse), as well as \emph{hyper-vulnerable} ones that have  exponential complexity. In addition and, most importantly, our technique can also construct an \emph{attack automaton} that captures all possible attack strings. The construction of attack automata is crucial for reasoning about input sanitization at the source-code level.

We have implemented the ideas proposed in this paper in a tool called \tool for finding vulnerabilities in Java applications. We use \tool to analyze dozens of web applications collected from Github repositories and show that \tool can detect 41 denial-of-service vulnerabilities with an 11\% false positive rate. Furthermore, we manually confirm these vulnerabilities and show that each of them can be exploited to cause the server to become unresponsive for at least 10 minutes.

To summarize, this paper makes the following  contributions:
\begin{itemize}
\item We present  algorithms for reasoning about 
worst-case complexity of NFAs.
Given an NFA $\automaton$, our algorithm can identify whether $\automaton$ has linear, super-linear, 
or exponential time complexity and can construct an \emph{attack automaton} that
accepts input strings that could cause worst-case behavior for $\automaton$.
\item We describe a program analysis  to automatically identify ReDoS vulnerabilities. Our technique
uses the results of the regular expression analysis to identify \emph{sinks} and reason about input sanitization using attack automata.
\item We use these ideas to build an end-to-end  tool  called \tool
for finding  vulnerabilities in Java. In our evaluation, we find 41 security vulnerabilities in 150 Java programs collected from Github with a 11\% false positive rate.

\end{itemize}

\section{Overview}
\label{sec:overview}

\begin{figure}[!t]
\begin{lstlisting}[style=java,xleftmargin=0.5cm]
public class RegExValidator {
  boolean validEmail(String t) { return t.matches(".+@.+\\.[a-z]+"); } ¤\label{re_suplin_1}¤
  boolean validComment(String t) {
    return !t.matches("(\\p{Blank}*(\\r?\\n)\\p{Blank}*)+"); } ¤\label{re_exp}¤
  boolean safeComment(String t) { return t.matches("([^\/<>])+"); } ¤\label{re_lin}¤
  boolean validUrl(String t) {
    return t.matches("www\\.shoppers\\.com/.+/.+/.+/.+/"); } ¤\label{re_suplin_2}¤
}
public class CommentFormValidator implements Validator {
  private Admin admin;
  public void validate(CommentForm form, Errors errors) { ¤\label{validate_begin}¤
    String senderEmail = form.getSenderEmail();
    String productUrl = form.getProductUrl();
    String comment = form.getComment();
    if (!RegExValidator.validEmail(admin.getEmail())) return; ¤\label{use_1}¤
    if (senderEmail.length() <= 254) { ¤\label{san_1}¤
      if (RegExValidator.validEmail(senderEmail)) ... } ¤\label{use_2}¤
    if (productUrl.split("/").length == 5) { ¤\label{san_3}¤
      if (RegExValidator.validUrl(productUrl)) ... } ¤\label{use_3}¤
    if (RegExValidator.safeComment(comment)) { ¤\label{san_2}¤
      if (RegExValidator.validComment(comment)) ... } ¤\label{use_4}¤
  } ¤\label{validate_end}¤
\end{lstlisting}
\vspace{-15pt}
\caption{\small Motivating example containing ReDoS vulnerabilities}
\vspace{-15pt}
\label{fig:overview-example}
\end{figure}

We illustrate our technique  using  the code snippet shown in Fig.~\ref{fig:overview-example}, which shows two relevant classes, namely {\tt
  RegExValidator}, that is used to validate that certain strings match a
given regular expression, and {\tt CommentFormValidator}, that checks the
validity of a comment form filled out by a user. In particular, the comment
form submitted by the user includes the user's email address, the URL of the
product about which the user wishes to submit a comment\footnote{Due to the store's organization, the URL is
  expected to be of the form\\\noindent{\tt \small
    www.shoppers.com/Dept/Category/Subcategory/product-id/}}, and the text containing the comment itself.
We now explain how our technique can  determine whether this program contains a denial-of-service vulnerability.

\begin{figure}
\begin{center}
\includegraphics[scale=0.18]{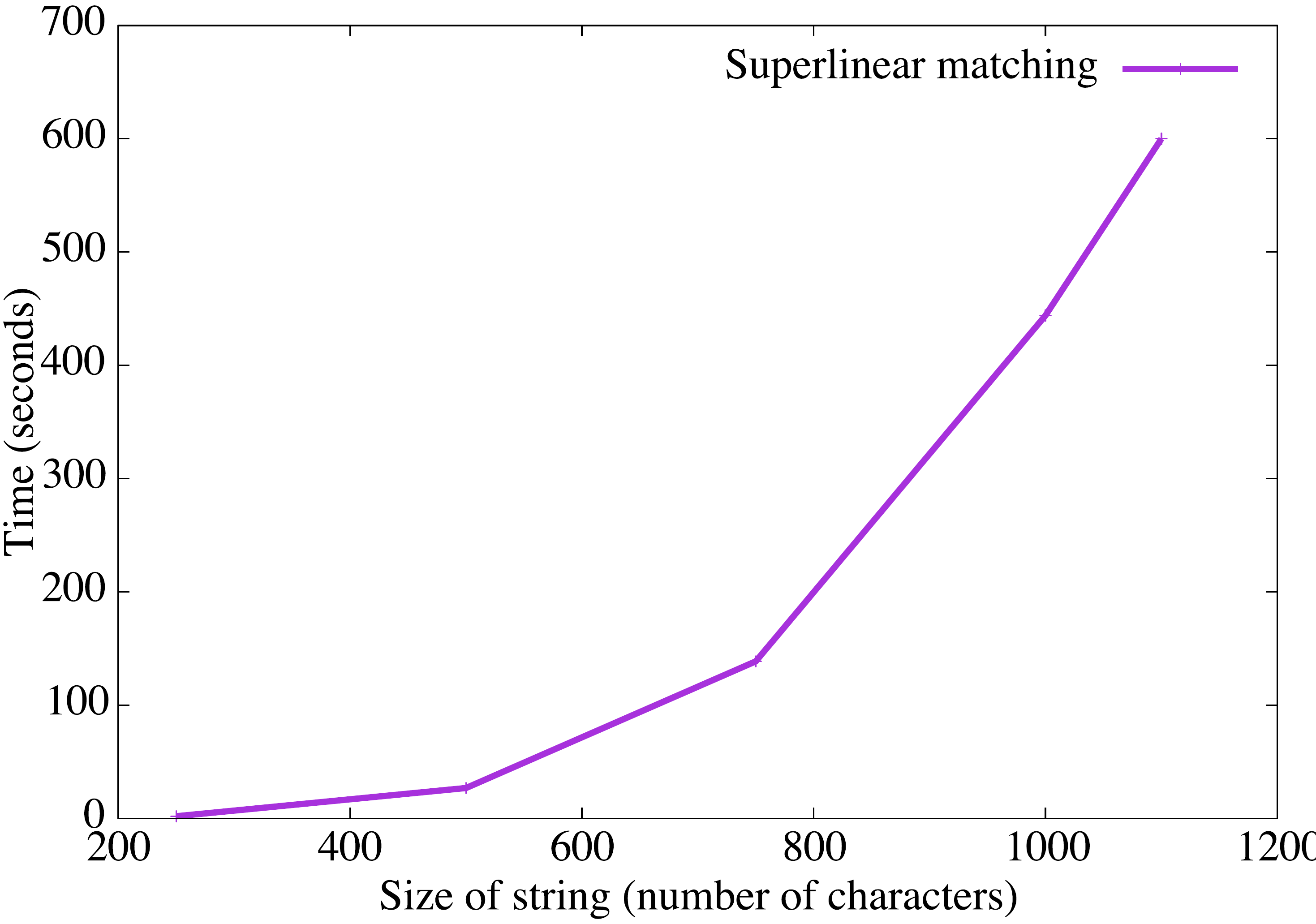}
\hfill
\includegraphics[scale=0.18]{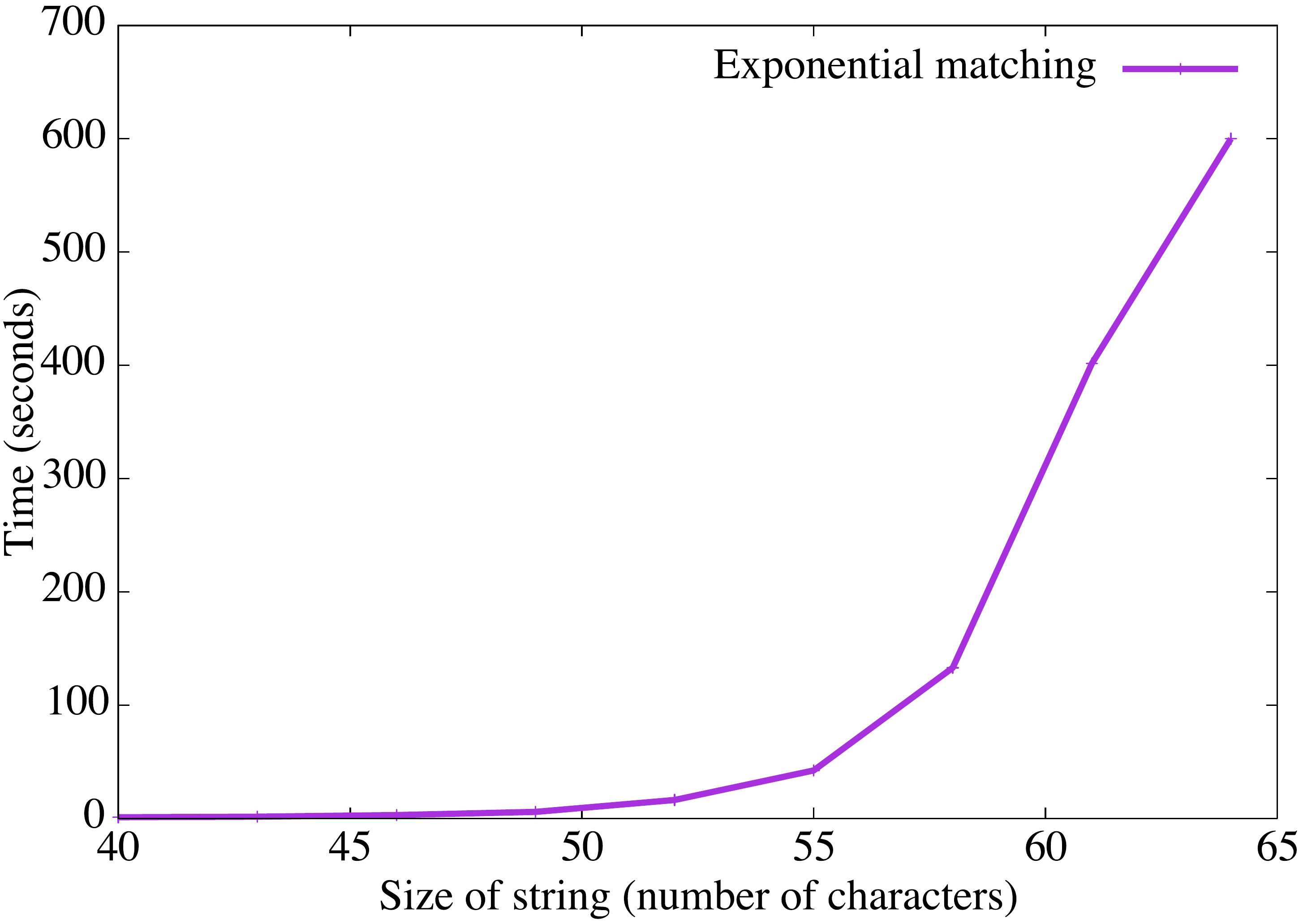}
\end{center}
\vspace{-0.2in}
\caption{\small Matching time against malicious string size for  vulnerable (left) and hyper-vulnerable (right) regular expressions from Fig.~\ref{fig:overview-example}.}
\label{fig:overview-plot}
\vspace{-0.2in}
\end{figure}

\vspace{0.1in} \noindent
{\bf \emph{Regular expression analysis.}}  For each regular expression  in the program, we construct its corresponding NFA and statically analyze it to determine whether its worst-case complexity is linear, super-linear, or exponential. For our running example, the NFA complexity analysis finds instances of each category. In particular, the regular expression used at line~\ref{re_lin} has linear matching complexity, while the one from  line~\ref{re_exp} has exponential  complexity. The regular expressions from lines~\ref{re_suplin_1} and~\ref{re_suplin_2} have super-linear (but not exponential) complexity.  Fig.~\ref{fig:overview-plot} plots input size against running time  for the regular expressions from lines~\ref{re_suplin_1} and~\ref{re_exp} respectively. For the super-linear and exponential regular expressions, our technique also constructs an attack automaton that recognizes all strings that cause worst-case behavior. In addition, for each regular expression, we determine a lower bound  on the length of any possible attack string using dynamic analysis.

\vspace{0.1in} \noindent
{\bf \emph{Program analysis.}} The presence of a vulnerable regular expression does not necessarily mean that the program itself is vulnerable. For instance, the vulnerable regular expression may not be matched against an attacker-controlled string, or the program may take measures to prevent the user from supplying a string that is an instance of the attack pattern. Hence, we also perform static analysis at the source code level to determine if the program is actually vulnerable.

Going back to our example, the {\tt validate} procedure (lines~\ref{validate_begin}--\ref{validate_end}) calls  {\tt validEmail}  to check whether the website administrator's email address is valid. Even though  {\tt validEmail}  contains a super-linear regular expression, line~\ref{use_1} does not contain a vulnerability because the administrator's email is not supplied by the user. Since our analysis tracks taint information, it does not report line~\ref{use_1} as being vulnerable.
Now, consider the second call to {\tt validEmail} at line~\ref{use_2}, which matches the vulnerable regular expression against user input. 
However, since the program bounds the size of the input string to be at most 254 (which is smaller than the lower bound identified by our analysis), line~\ref{use_2} is also not vulnerable.

Next, consider  the call to {\tt validUrl} at  line~\ref{use_3}, where  {\tt productUrl} is a user input.  At first glance, this  appears to be a vulnerability because  the matching time of the regular expression from line~\ref{re_exp} against a malicious input string
grows quite rapidly with input size (see Fig.~\ref{fig:overview-plot}). However, the check at line~\ref{san_3} actually prevents calling {\tt validUrl} with an attack string:  Specifically, our  analysis determines that attack strings must be of the form 
{\tt www.shoppers.com}$\cdot${\tt/}$^b\cdot${\tt/}$^+\cdot${\tt x},
where {\tt x} denotes any character and  $b$ is a constant inferred by our analysis (in this case, much greater than $5$). Since our program analysis  also reasons about input sanitization, it can establish that line~\ref{use_3} is safe.

Finally, consider the call to {\tt validComment} at line~\ref{use_4}, where {\tt comment} is again a user input and is matched against a regular expression with exponential complexity. Now, the question is whether the condition at line~\ref{san_2} prevents  {\tt comment} from conforming to the attack pattern \verb+\n\t\n\t+$($ \verb+\t\n\t+$) ^k${\tt a}. Since this is not the case,  line~\ref{use_4} actually contains a serious DoS vulnerability.

\vspace{0.1in} \noindent
{\bf \emph{Summary of challenges.}} This example illustrates several challenges we must address:  First, given a regular expression $\regex$, we must  reason about the worst-case time complexity of its corresponding NFA. Second, given vulnerable regular expression $\regex$, we must determine whether the program allows  $\regex$ to be matched against a string that is (a) controlled by the user,  (b) is an instance of the attack pattern for  regular expression $\regex$, and (c) is large enough to cause the matching algorithm to take significant time.

Our approach solves these challenges by combining complexity analysis of NFAs with sanitization-aware taint analysis. The key idea that makes this combination possible is to produce an attack automaton for each vulnerable NFA. Without such an attack automaton, the program analyzer cannot effectively determine whether an input string can correspond to an attack string.

\begin{wrapfigure}{r}{0.55 \textwidth}
  \vspace{-0.3in}
  \begin{center}
\includegraphics[scale=0.17]{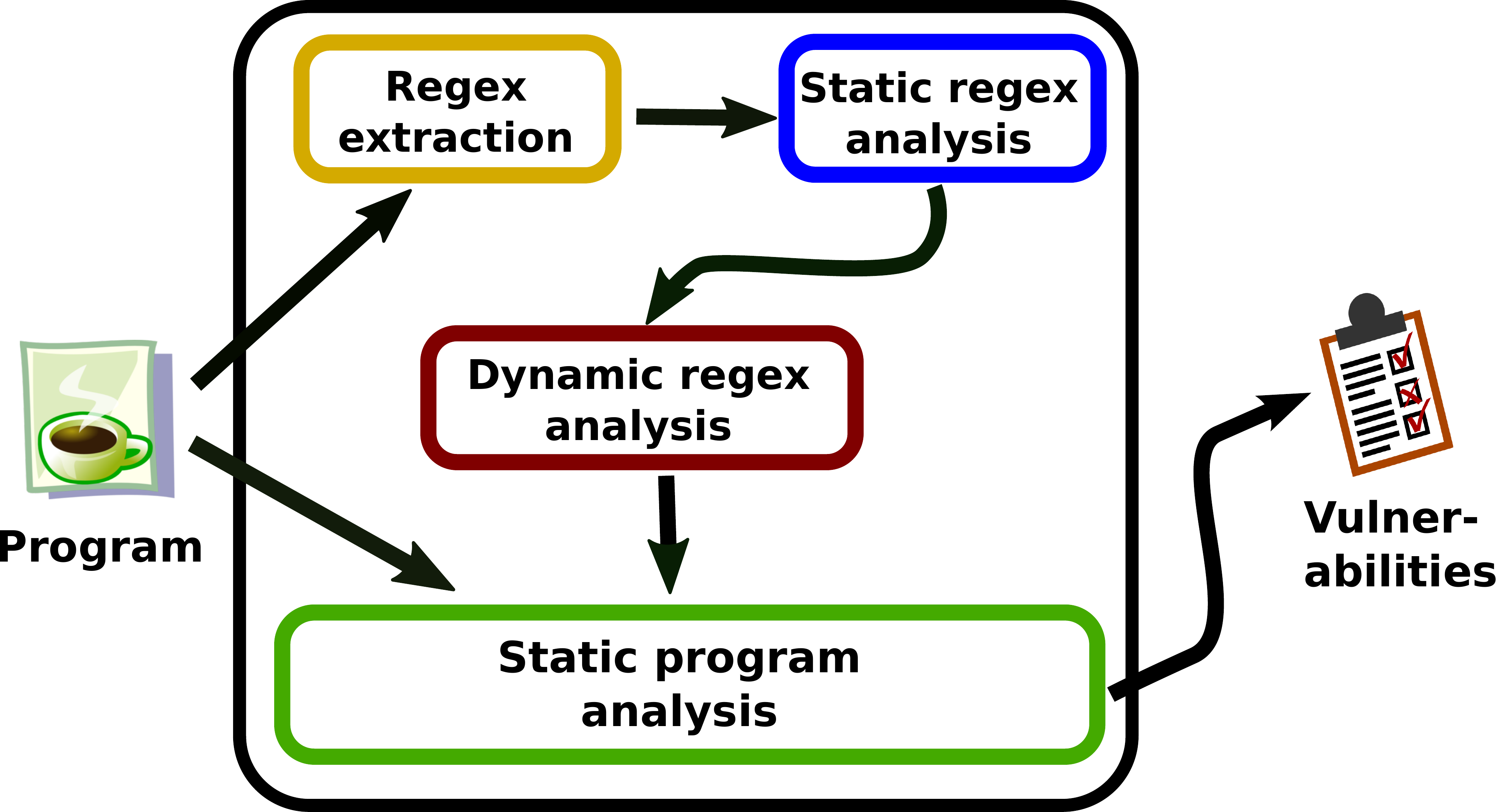}
  \end{center}
  \vspace{-0.2in}
  \caption{Overview of our approach}
  \label{fig:architecture}
  \vspace{-0.2in}
\end{wrapfigure}


As shown in Fig.~\ref{fig:architecture}, the \tool toolchain incorporates both static and dynamic regular expression analysis. 
The static analysis creates attack patterns $s_0 \cdot s^k \cdot s_1$ and dynamic analysis infers a lower bound $b$ on the number of occurrences of $s$ in order to exceed a minimum runtime threshold. The program analysis uses both the attack automaton and the lower bound $b$ to  reason about  input sanitization.


\section{Preliminaries}\label{sec:prelim}
This section presents some useful background and terminology.

\begin{definition}{\bf (NFA)} An NFA $\automaton$ is a $5$-tuple $(Q, \Sigma, \Delta, q_0, F)$ where $Q$ is a finite set of states, $\Sigma$ is a finite alphabet of symbols, and $\Delta: Q \times \Sigma \rightarrow 2^Q$ is the transition function. Here, $q_0 \in Q$ is the \emph{initial state}, and $F \subseteq Q$ is the set of \emph{accepting states}. We say that $(q, l, q')$ is a transition via label $l$ if $q' \in \Delta(q, l)$. 
\end{definition}

An NFA $\automaton$ accepts a string $s = a_0a_1\ldots a_n$ iff there exists a sequence of states $q_0, q_1, ..., q_n$ such that $q_n \in F$ and $q_{i+1} \in \Delta(q_i, a_i)$. The language of $\automaton$, denoted $\langu(\automaton)$, is the set of all strings that are accepted by $\automaton$.
Conversion from a regular expression to an NFA is sometimes referred to as \emph{compilation} and can be achieved using well-known techniques, such as Thompson's  algorithm~\cite{thompson}. 

In this paper, we assume that  membership  in a  regular language $\langu(\regex)$ is decided through a worst-case exponential algorithm that performs backtracking search over  possible runs of the NFA representing $\regex$.  
While there exist linear-time matching algorithms (e.g., based on DFAs), many real-world libraries employ backtracking search for two key reasons: First, the compilation of a regular expression is much faster using NFAs and uses much less memory (DFA's can be exponentially larger). Second, the backtracking search approach can handle regular expressions  containing extra features like backreferences and lookarounds. Thus, many widely-used libraries (e.g., {\tt java.util.regex}, Python's standard library) employ backtracking search for  regular expression matching.

In the remainder of this paper, we will use the notation $\automaton^*$ and $\automaton^\emptyset$ to denote the NFA that accepts $\Sigma^*$ and the empty language respectively. Given two NFAs $\automaton_1$ and $\automaton_2$, we write $\automaton_1 \cap \automaton_2$, $\automaton_1 \cup \automaton_2$, and $\automaton_1 \cdot \automaton_2$ to denote automata intersection, union, and concatenation. 
Finally, given an automaton $\automaton$, we  write $\overline{\automaton}$ to represent its complement, and we use the notation $\automaton^+$ to represent the NFA that recognizes exactly the language $\{ s^k \ | \ k \geq 1 \land s \in \langu(\automaton) \}$.


\begin{definition}{\bf (Path)}
\label{def:path}
Given an NFA $\automaton = (Q, \Sigma, \Delta, q_0, F)$, a path $\npath$ of $\automaton$ is a sequence of transitions 
$ (q_1, \labl_1, q_2), \ldots, (q_{m-1},  \labl_{m-1}, q_m) $
where $q_i \in Q$,  $\labl_i \in \Sigma$, and $q_{i+1} \in  \Delta(q_{i}, \labl_i )$. We say that $\npath$ starts in $q_i$ and ends at $q_{m}$, and we write
$\labels(\npath)$ to denote the sequence of labels $(\labl_1, \ldots, \labl_{m-1})$. 
\end{definition}



\section{Detecting Hyper-Vulnerable NFAs}
\label{sec:regex}

In this section, we explain our technique for determining if an NFA is \emph{hyper-vulnerable} and
show how to generate an \emph{attack automaton} that recognizes exactly the set of attack strings.


\begin{definition}{\bf (Hyper-Vulnerable NFA)}
An NFA $\automaton = (Q, \Sigma, \Delta, q_0, F)$ is hyper-vulnerable iff there exists a
backtracking search algorithm $\textsc{Match}$ over the paths of $\automaton$ such that the worst-case complexity  of $\textsc{Match}$
is exponential in the length of the input string.
\end{definition}

We will demonstrate that an NFA $\automaton$ is hyper-vulnerable by showing that there exists a string $s$ such 
that the number of distinct matching paths $\npath_i$ from state $q_{0}$ to a  rejecting state $q_r$ with
$labels(\npath_i) = s$ is exponential in the length of $s$. Clearly, if $s$ is rejected by $\automaton$, then $\textsc{Match}$ will need to explore each of these exponentially many paths. Furthermore,  even if $s$ is accepted by $\automaton$, there
exists a backtracking search algorithm (namely, the one that explores all rejecting paths first) that results in exponential worst-case behavior.

\begin{theorem}
An NFA $\automaton = (Q, \Sigma, \Delta, q_0, F)$ is \emph{hyper-vulnerable} iff
  there exists a \emph{pivot state} $q \in Q$ and two distinct paths $\npath_1, \npath_2$ such that
  (i) both $\npath_1, \npath_2$ start and end at $q$, (ii) $\labels(\npath_1)
  = \labels(\npath_2)$, and (iii) there is a path $\npath_p$ from initial
  state $q_0$ to $q$, and (iv) there is a path $\npath_s$ from $q$ to a state $q_r
  \not \in F$.
\label{thm:hyper-vulnerable}
\end{theorem}
\begin{proof}
\ifdefined\EXTENDEDVERSION
The sufficiency argument is laid out below, and the necessity argument can be found in the appendix. 
\else
The sufficiency argument is laid out below, and the necessity argument can be
found in the extended version of this paper~\cite{WuestholzOlivo2017}. 
\fi
\end{proof}

\begin{wrapfigure}{r}{0.5 \textwidth}
\vspace{-0.5in}
\begin{center}
\begin{tikzpicture}[->,>=stealth, photon/.style={decorate,decoration={snake,post length=1mm}}]
\node [circle] (a) at (0,0) {\large $q_0$};
\node [circle] (b) at (2,0) {\large $q$};
\node (e) at (2,-0.5) {\textcolor{blue}{pivot}};
\node [circle] (c) at (4,0) {\large $q_r$};
\node (f) at (4.3,1) {$\emph{labels}(\pi_1){=}\emph{labels}(\pi_2)$};
\draw[thick,photon] (a) -- node [below] {$\pi_p$} (b);
\draw[thick,photon] (a) -- node [above] {\textcolor{blue}{prefix~~}} (b);
\draw[thick,photon] (b) -- node [above] {\textcolor{blue}{\phantom{p}~suffix}} (c);
\draw[thick,photon] (b) -- node [below] {$\pi_s$} (c);
\draw[thick] (b) edge [loop below,looseness=8,snake it,in=150,out=30] node {$\pi_1$} (b);
\draw[thick] (b) edge [loop below,looseness=8,snake it,in=330,out=210] node {$\pi_2$} (b);
\end{tikzpicture}
\end{center}
\vspace{-0.3in}
\caption{Hyper-vulnerable NFA pattern
\vspace{-0.2in}
}
\label{fig:exponential}
\end{wrapfigure}
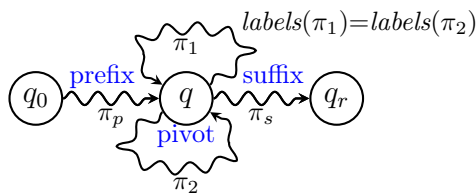

To gain intuition about hyper-vulnerable NFAs,  consider Fig.~\ref{fig:exponential} illustrating the conditions of Theorem~\ref{thm:hyper-vulnerable}. First, a hyper-vulnerable NFA must contain a \emph{pivot state} $q$, such that, starting at $q$, there are two different ways (namely, $\pi_1, \pi_2$) of getting back to $q$ on the same input string $s$ (i.e., $\emph{labels}(\pi_1)$). Second, the pivot state $q$ should be reachable from the initial state $q_0$, and there must be a way of reaching a rejecting state $q_r$ from $q$.

To understand why these conditions cause exponential behavior,  consider a string of the form $s_0 \cdot s^k \cdot s_1$, where
$s_0$ is the \emph{attack prefix} given by $\emph{labels}(\pi_p)$, $s_1$ is the \emph{attack suffix} given by $\emph{labels}(\pi_s)$, and $s$ is the \emph{attack core} given by $\emph{labels}(\pi_1)$.
Clearly, there is an execution path of $\automaton$ in which the string $s_0
\cdot s^k \cdot s_1$ will be rejected. For example, $\pi_p \cdot \pi_1^k \cdot
\pi_s$ is exactly such a path.

  \begin{wrapfigure}{l}{0.58\textwidth}
\ifdefined\EXTENDEDVERSION
  \vspace{-0.6in}
\else
  \vspace{-0.3in}
\fi
    \begin{minipage}{0.58\textwidth}
      \begin{algorithm}[H]
        \caption{Hyper-vulnerable NFA}
        \label{Alg:algo1}
        \begin{algorithmic}[1]
  \Function{AttackAutomaton}{$\automaton$}
\State assume $\automaton = (Q, \Sigma, \Delta, q_0, F)$
\State $\attack \gets \automaton^{\emptyset}$
\For{$q_i \in Q$}
\State $\attack_i \gets \Call{AttackForPivot}{\automaton, q_i}$
\State $\attack \gets \attack \cup \attack_i$
\EndFor
\State \Return $\attack$
\EndFunction
\Function{AttackForPivot}{$\automaton, q$}
\State assume $\automaton = (Q, \Sigma, \Delta, q_0, F)$
\State $\attack \gets \automaton^{\emptyset}$ \label{line:init}
\For{$(q, l, q_1), (q, l, q_2) \in \Delta \wedge q_1 \neq q_2$}\label{line:for_all_states}
\State $\automaton_1 \gets \Call{LoopBack}{\automaton, q, l, q_1}$\label{line:loop-back-1}
\State $\automaton_2 \gets \Call{LoopBack}{\automaton, q, l, q_2}$\label{line:loop-back-2}
\State $\automaton_p \gets (Q, \Sigma, \Delta, q_0, \{q\})$\label{line:prefix}
\State $\automaton_s \gets (Q, \Sigma, \Delta, q, F)$\label{line:suffix}
\State $\attack \gets \attack \cup (\automaton_p \cdot (\automaton_1 \cap \automaton_2)^+ \cdot \overline{\automaton_s})$
\EndFor
\State \Return $\attack$
\EndFunction
\Function{LoopBack}{$\automaton, q, l, q'$} \label{line:loop-back-start}
\State assume $\automaton = (Q, \Sigma, \Delta, q_0, F)$
\State $q^\star \gets \Call{NewState}{Q}$
\State $Q' \gets Q \cup q^\star $; \ \ $\Delta' \gets \Delta \cup (q^\star, l, q')$
\State \Return $(Q', \Sigma, \Delta', q^\star, \{ q \})$ \label{line:loop-back-end}
\EndFunction
        \end{algorithmic}
      \end{algorithm}
    \end{minipage}
    \vspace{-0.3in}
  \end{wrapfigure}

Now, consider a string $s_0 \cdot s^{k+1} \cdot s_1$ that has an additional instance of the attack core $s$ in the middle, and suppose that there are $n$ possible executions of $\automaton$ on the prefix $s_0 \cdot s^k$ that  end in $q$. Now, for each of these $n$ executions, there are two ways to get back to $q$ after reading $s$:  one that takes path $\pi_1$ and another that takes path $\pi_2$. Therefore, there are $2n$ possible executions of $\automaton$ that end in $q$. Furthermore, the matching algorithm will (in the worst case) end up exploring all of these $2n$ executions since there is a way to reach the rejecting state $q_r$. Hence, we end up doubling the running time of the algorithm every time we add an instance of the attack core $s$ to the middle of the input string.

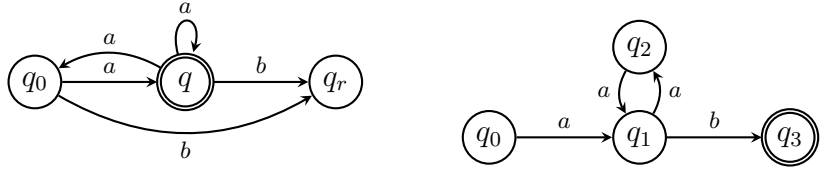
\begin{figure}[t]
~~~~~
\begin{tikzpicture}[->,>=stealth]
\node [circle] (a) at (0,0) {\large $q_0$};
\node [circle,accepting] (b) at (2,0) {\large $q$};
\node [circle] (c) at (4,0) {\large $q_r$};
\draw[thick] (a) edge node [above] {$a$} (b);
\draw[thick] (b) edge [bend right] node [above] {$a$} (a);
\draw[thick] (a) edge [bend right] node [below] {$b$} (c);
\draw[thick] (b) edge node [above] {$b$} (c);
\draw[thick] (b) edge [loop above] node {$a$} (b);
\end{tikzpicture}
\hfill
\begin{tikzpicture}[->,>=stealth]
\node [circle] (a) at (0,0) {\large $q_0$};
\node [circle] (b) at (2,0) {\large $q_1$};
\node [circle] (d) at (2,1.2) {\large $q_2$};
\node [circle,accepting] (c) at (4,0) {\large $q_3$};
\draw[thick] (a) edge node [above] {$a$} (b);
\draw[thick] (b) edge node [above] {$b$} (c);
\draw[thick] (b) edge [bend right] node [right] {$a$} (d);
\draw[thick] (d) edge [bend right] node [left] {$a$} (b);
\end{tikzpicture}
~~~~~
\vspace{-0.1in}
\caption{A hyper-vulnerable NFA (left) and an attack automaton (right).}\label{fig:ex-vulnerable}
\vspace{-0.1in}
\end{figure}

\vspace{-5pt}
\begin{example}\label{ex:vulnerable}
  The NFA in Fig.~\ref{fig:ex-vulnerable} (left) is hyper-vulnerable because there exist two different paths $\pi_1 = (q, a, q), (q, a, q)$ and  $\pi_2 = (q, a, q_0), (q_0, a, q)$ that contain the same labels and that start and end in $q$.
Also, $q$ is reachable from $q_0$, and the rejecting state $q_r$ is
reachable from $q$. Attack strings for this NFA are of the form $a
\cdot (a \cdot a)^k \cdot b$, and the attack automaton is shown in Fig.~\ref{fig:ex-vulnerable} (right).
\end{example}

We now use Theorem~\ref{thm:hyper-vulnerable} to devise Algorithm~\ref{Alg:algo1} for constructing the attack automaton $\attack$ for a given NFA.
The key idea of our algorithm is to search for all possible pivot states $q_i$  and construct the attack automaton $\attack_i$ for state $q_i$. 
The full attack automaton  is then obtained as the union of all  $\attack_i$. 
Note that Algorithm~\ref{Alg:algo1} can be used to determine if automaton $\automaton$ is vulnerable: $\automaton$ exhibits 
worst-case exponential behavior iff the language accepted by $\attack$ is non-empty.

In Algorithm~\ref{Alg:algo1}, most of the real work is done by the {\sc AttackForPivot} procedure, which constructs the attack automaton for a specific state $q$: Given a pivot state $q$, we want to find two different paths $\pi_1$, $\pi_2$ that loop back to $q$ and that have the same set of labels.  Towards this goal, line~\ref{line:for_all_states} of Algorithm~\ref{Alg:algo1} considers all pairs of transitions from $q$ that have the same label (since we must have $\emph{labels}(\pi_1) = \emph{labels}(\pi_2)$).

Now, let us consider a pair of transitions $\tau_1 = (q, l, q_1)$ and $\tau_2 = (q, l, q_2)$. For each $q_i$ ($i \in \{1,2\})$, we want to find all strings that start in $q$, take transition $\tau_i$, and then loop back to $q$. In order to find all such strings $\mathcal S$, Algorithm~\ref{Alg:algo1} invokes the {\sc LoopBack} function (lines~\ref{line:loop-back-start}--\ref{line:loop-back-end}), which constructs an automaton $\automaton'$ that recognizes exactly $\mathcal S$. Specifically,  the final state of  $\automaton'$  is  $ q $ because we want to loop back to state $q$. Furthermore, $\automaton'$ contains a new initial state $q^*$ (where $q^* \not \in Q$) and a single outgoing transition $(q^*, l, q_i)$ out of $q^*$ because we only want to consider paths that take the transition to $q_i$ first. Hence, each $\automaton_i$ in lines~\ref{line:loop-back-1}--\ref{line:loop-back-2} of the {\sc AttackForPivot} procedure corresponds to a set of paths that loop back to $q$ through state $q_i$. Observe that, if a string $s$ is accepted by $\automaton_1 \cap \automaton_2$, then $s$ is an attack core for pivot state $q$.

We now turn to the problem of computing the set of all attack prefixes and suffixes for pivot state $q$: In line~\ref{line:prefix} of Algorithm~\ref{Alg:algo1},  $\automaton_p$ is the same as the original NFA $\automaton$ except that its only accepting state is $q$. Hence, $\automaton_p$ accepts all attack prefixes for pivot $q$. Similarly, $A_s$ is the same as $\automaton$ except that its initial state is $q$ instead of $q_0$; thus, $\overline{A_s}$ accepts all attack suffixes for $q$.

Finally, let us consider how to construct the full attack automaton  $\attack$ for $q$. As explained earlier, all attack strings are of the form $s_1 \cdot s^k \cdot s_2$ where $s_1$ is the attack prefix, $s$ is the attack core, and $s_2$ is the attack suffix. Since  $\automaton_p$, $\automaton_1 \cap \automaton_2$, and $\overline{\automaton_s}$ recognize attack prefixes, cores, and suffixes respectively, any string that is accepted by $\automaton_p \cdot (\automaton_1 \cap \automaton_2)^+ \cdot \overline{\automaton_s}$ is an attack string for the original NFA $\automaton$.

\begin{theorem}{\bf (Correctness of Algorithm~\ref{Alg:algo1})}\ifdefined\EXTENDEDVERSION\footnote{The proofs of Theorems~\ref{thm:algo1} and~\ref{thm:algo2} are given in the appendix.}\else\footnote{The proofs of Theorems~\ref{thm:algo1} and~\ref{thm:algo2} are given in the extended version of this paper~\cite{WuestholzOlivo2017}.}\fi\label{thm:algo1}
Let  $\attack$ be the result of calling $\textsc{AttackAutomaton}(\automaton)$
for NFA $\automaton = (Q, \Sigma, \Delta, q_0, F)$. 
  For every $s \in \langu(\attack)$, there exists a rejecting state $q_r \in Q \setminus F$ s.t. the
  number of distinct paths $\npath_i$ from $q_0$ to $q_r$ with
  $labels(\npath_i) = s$ is exponential in the number of repetitions of the attack core in $s$.
\end{theorem}

\vspace{-15pt}

\section{Detecting Vulnerable NFAs}
\label{sec:regex-superlinear}
So far, we only considered the problem of identifying NFAs whose worst-case running time is exponential. However, in practice, even NFAs with super-linear complexity can cause catastrophic backtracking. In fact, many acknowledged ReDoS vulnerabilities (e.g.,~\cite{cve-superlinear1,cve-superlinear2,cve-superlinear3}) involve regular expressions whose matching complexity is ``only"  quadratic.
Based on this observation, we extend the techniques from the previous section to statically detect NFAs with super-linear time complexity. Our solution builds on insights from Section~\ref{sec:regex} to
construct an attack automaton for this larger class of vulnerable regular expressions.

\subsection{Understanding Super-Linear NFAs}
\label{subsec:vulnerable-nfa-super-linear}

Before we present the algorithm for detecting super-linear NFAs, we
provide a theorem that explains the correctness of our solution.


\begin{definition}{\bf (Vulnerable NFA)}
An NFA $\automaton = (Q, \Sigma, \Delta, q_0, F)$ is vulnerable iff there exists a
backtracking search algorithm $\textsc{Match}$ over the paths of $\automaton$ such that the worst-case complexity  of $\textsc{Match}$
is at least quadratic in the length of the input string.
\end{definition}

\begin{theorem}
 An NFA $\automaton= (Q, \Sigma, \Delta, q_0, F)$ is \emph{vulnerable} iff
  there exist two states $q \in Q$ (\emph{the pivot}), $q' \in Q$,
 and three paths $\npath_1$,
  $\npath_2$, and $\npath_3$ (where $\npath_1 \neq \npath_2$) such that
  (i) $\npath_1$ starts and ends at $q$, (ii) $\npath_2$ starts at $q$ and ends
  at $q'$, (iii) $\npath_3$ starts and ends
  at $q'$, (iv) $\labels(\npath_1)
  = \labels(\npath_2) = \labels(\npath_3)$, and (v) there is a path $\npath_p$ from  $q_0$ to $q$,  (vi) there is a path $\npath_s$ from $q'$ to a state $q_r
  \not \in F$.
\label{thm:vulnerable-super-linear}
\end{theorem}

\begin{proof}
\ifdefined\EXTENDEDVERSION
The necessity argument can be found in the appendix. The sufficiency argument is in the following text.
\else
The sufficiency argument is laid out below, and the necessity argument can be found in the extended version of this paper~\cite{WuestholzOlivo2017}.
\fi

\end{proof}

\begin{figure}[t]
\begin{center}
\begin{tikzpicture}[->,>=stealth, photon/.style={decorate,decoration={snake,post length=1mm}}]
\node [circle] (a) at (0,0) {\large $q_0$};
\node [circle] (b) at (2,0) {\large $q$};
\node [circle] (d) at (4,0) {\large $q'$};
\node (e) at (2,-0.5) {\textcolor{blue}{pivot}};
\node [circle] (c) at (6,0) {\large $q_r$};
\node (f) at (6.3,1) {$\emph{labels}(\pi_1) = \emph{labels}(\pi_2)$};
\node (f) at (6.3,0.67) {$\phantom{\emph{labels}(\pi_1)} = \emph{labels}(\pi_3)$};
\draw[thick,photon] (a) -- node [below] {$\pi_p$} (b);
\draw[thick,photon] (b) -- node [below] {$\pi_2$} (d);
\draw[thick,photon] (a) -- node [above] {\textcolor{blue}{prefix~~}} (b);
\draw[thick,photon] (d) -- node [above] {\textcolor{blue}{\phantom{p}~suffix}} (c);
\draw[thick,photon] (d) -- node [below] {$\pi_s$} (c);
\draw[thick] (b) edge [loop below,looseness=8,snake it,in=150,out=30] node {$\pi_1$} (b);
\draw[thick] (d) edge [loop below,looseness=8,snake it,in=150,out=30] node {$\pi_3$} (d);
\end{tikzpicture}
\end{center}
\vspace{-0.3in}
\caption{General pattern characterizing vulnerable NFAs
}
\label{fig:super-linear}
\vspace{-0.1in}
\end{figure}
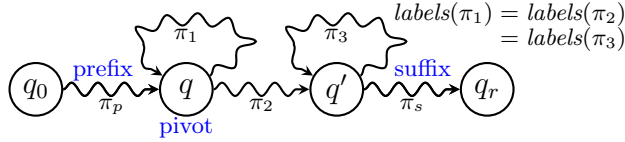

Fig.~\ref{fig:super-linear} illustrates the intuition behind the conditions
above. The distinguishing characteristic of a super-linear NFA is that it contains
two states $q, q'$ such that $q'$ is reachable from $q$
on input string $s$, and  it is possible to loop back from $q$ and $q'$ to the same state on string $s$.
 In addition, just like in Theorem~\ref{thm:hyper-vulnerable},
the pivot state $q$ needs to be reachable
from the initial state, and a rejecting state $q_r$ must be reachable from
$q'$.
Observe that any automaton that is hyper-vulnerable according to
Theorem~\ref{thm:hyper-vulnerable} is also vulnerable according to Theorem~\ref{thm:vulnerable-super-linear}. Specifically, consider an automaton $\automaton$  with two distinct paths  $\pi_1, \pi_2$
that loop around $q$. In this case, if we take $q'$ to be $q$ and $\pi_3$ to be $\pi_1$, we immediately see
that $\automaton$ also satisfies the conditions of Theorem~\ref{thm:vulnerable-super-linear}.

To understand why the conditions of Theorem~\ref{thm:vulnerable-super-linear} imply
 super-linear time complexity, let us consider a  string  of the form $s_0 \cdot s^k \cdot s_1$
where  $s_0$ is the \emph{attack prefix} given by $\emph{labels}(\pi_p)$, $s_1$ is the \emph{attack suffix} given by $\emph{labels}(\pi_s)$, and $s$ is the \emph{attack core} given by $\emph{labels}(\pi_1)$. Just like in the previous section, the path $\pi_p \,\pi_1^k\, \pi_s$
describes an execution for rejecting the string $s_0 \cdot s^k \cdot s_1$ in
automaton $\automaton$. Now, let $T_q(k)$ represent the running time of rejecting the
string $s^k s_1$ starting from $q$, and suppose that it takes $1$ unit of time to
read string $s$. We can write the following recurrence relation for $T_q(k)$:
$$
T_q(k) = (1 + T_q(k-1)) + (1 + T_{q'}(k-1))
$$
To understand where this recurrence is coming from, observe that there are two ways to process the first occurence of  $s$:
\begin{itemize}
\item Take path $\pi_1$ and come back to $q$, consuming 1 unit of time to process string $s$. Since we are back at $q$, we still have $T_q(k-1)$ units of work to perform.
\item Take path $\pi_2$ and proceed to $q'$, also consuming 1 unit of time to process string $s$. Since we are now at $q'$, we  have $T_{q'}(k-1)$ units of work to perform.
\end{itemize}

Now, observe that a lower bound on $T_{q'}(k)$ is $k$ since one way to reach $q_r$ is $\pi_3^k \pi_s$, which requires us to read the entire input string. This observation allows us to obtain the following recurrence relation:
$$
T_q(k) \geq T_q(k-1) + k + 1
$$
Thus, the running time of $\automaton$ on the input string $s_0 \cdot s^k \cdot s_1$ is at least $k^2$.

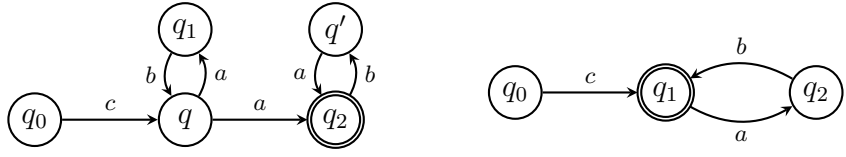
\begin{figure}[t]
~~~~
\begin{tikzpicture}[->,>=stealth]
\node [circle] (a) at (0,0) {\large $q_0$};
\node [circle] (b) at (2,0) {\large $q$};
\node [circle] (d) at (2,1.2) {\large $q_1$};
\node [circle,accepting] (c) at (4,0) {\large $q_2$};
\node [circle] (f) at (4,1.2) {\large $q'$};
\draw[thick] (a) edge node [above] {$c$} (b);
\draw[thick] (b) edge node [above] {$a$} (c);
\draw[thick] (b) edge [bend right] node [right] {$a$} (d);
\draw[thick] (d) edge [bend right] node [left] {$b$} (b);
\draw[thick] (c) edge [bend right] node [right] {$b$} (f);
\draw[thick] (f) edge [bend right] node [left] {$a$} (c);
\end{tikzpicture}
\hfill
\begin{tikzpicture}[->,>=stealth]
\node [circle] (a) at (0,0) {\large $q_0$};
\node [circle,accepting] (b) at (2,0) {\large $q_1$};
\node [circle] (c) at (4,0) {\large $q_2$};
\draw[thick] (a) edge node [above] {$c$} (b);
\draw[thick] (b) edge [bend right] node [below] {$a$} (c);
\draw[thick] (c) edge [bend right] node [above] {$b$} (b);
\end{tikzpicture}
~~~~
\caption{A vulnerable NFA (left) and its attack automaton (right).}
\label{fig:super-linear-ex}
\end{figure}


\begin{example}\label{ex:super-linear}
The NFA shown in Fig.~\ref{fig:super-linear-ex} (left) exhibits super-linear complexity because we can get from $q$ to $q'$ on input string $ab$, and  for both $q$ and $q'$, we loop back to the same state when reading input string $ab$. Specifically, we have:
$$
\begin{array}{lll}
\pi_1:  (q, a, q_1), (q_1, b, q) \ \ \ \ & 
\pi_2:  (q, a, q_2), (q_2, b, q') \ \ \ \ & 
\pi_3:  (q', a, q_2), (q_2, b, q')
\end{array}
$$
Furthermore, $q$ is reachable from $q_0$, and there exists a rejecting state, namely $q'$ itself, that is reachable from $q'$. The attack strings are of the form $c (ab)^k$, and Fig.~\ref{fig:super-linear-ex} (right) shows the attack automaton.
\end{example}


\begin{algorithm}[!]
\caption{Construct  super-linear attack automaton $\attack$ for  $\automaton$ and
  pivot~$q$}
\label{Alg:algo2}
\begin{algorithmic}[1]
\Function{AnyLoopBack}{$\automaton, q'$}
\State assume $\automaton = (Q, \Sigma, \Delta, q_0, F)$
\State $q^\star \gets \Call{NewState}{Q}$; \ \ $Q' \gets Q \cup q^\star $; \ \ $\Delta' \gets \Delta$
\For{$(q', l, q_i) \in \Delta$} \label{line:any_transition_1}
\State $\Delta' \gets \Delta' \cup (q^\star, l, q_i)$ \label{line:any_transition_2}
\EndFor
\State $\automaton' \gets (Q', \Sigma, \Delta', q^\star, \{ q' \})$
\State \Return $\automaton'$
\EndFunction
\Statex \vspace{-10pt}
\Function{AttackForPivot}{$\automaton, q$}
\State assume $\automaton = (Q, \Sigma, \Delta, q_0, F)$
\State $\attack \gets \automaton^{\emptyset}$
\For{$(q, l, q_1) \in \Delta \wedge (q, l, q_2) \in \Delta \wedge q_1 \neq q_2$} \label{line:for_transitions}
\State $\automaton_1 \gets \Call{LoopBack}{\automaton, q, l, q_1}$ \label{line:A1}
\State $\automaton_p \gets (Q, \Sigma, \Delta, q_0, \{ q \})$ \label{line:Ap}
\For{$q' \in Q$}
\State $q_i \gets \Call{NewState}{Q}$ \label{line:forQ}
\State $\automaton_2 \gets (Q \cup \{q_i\}, \Sigma, \Delta \cup \{(q_i, l,
q_2)\}, q_i, \{ q' \})$ \label{line:A2}
\State $\automaton_3 \gets \Call{AnyLoopBack}{\automaton, q'}$ \label{line:A3}
\State $\automaton_s \gets (Q, \Sigma, \Delta, q', F)$ \label{line:As}
\State $\attack \gets \attack \cup (\automaton_p \cdot (\automaton_1 \cap \automaton_2 \cap \automaton_3)^+ \cdot \overline{\automaton_s})$ \label{line:Aa}
\EndFor
\EndFor
\State \Return $\attack$
\EndFunction
\end{algorithmic}
\end{algorithm}

\vspace{-15pt}

\subsection{Algorithm for Detecting Vulnerable NFAs}

Based on the observations from the previous subsection, we can now formulate an algorithm that constructs
an attack automaton $\attack$ for a given automaton $\automaton$. Just like in
Algorithm~\ref{Alg:algo1}, we construct an attack automaton $\attack_i$ for
each state in $\automaton$ by invoking the \textsc{AttackForPivot} procedure. We then take
the union of all such $\attack_i$'s to obtain an automaton $\attack$ whose language
consists of strings that cause super-linear running time for $\automaton$.

Algorithm~\ref{Alg:algo2} describes the \textsc{AttackForPivot} procedure for
the super-linear case.
Just like in Algorithm~\ref{Alg:algo1}, we  consider all pairs of
 transitions from $q$ with the same label  (line~\ref{line:for_transitions}).
Furthermore, as in Algorithm~\ref{Alg:algo1}, we construct an automaton $\automaton_p$ that recognizes
attack prefixes for $q$ (line~\ref{line:Ap}) as well as an automaton $\automaton_1$ that recognizes non-empty strings that start and end at $q$ (line~\ref{line:A1}).

The key difference of Algorithm~\ref{Alg:algo2}  is that we also need to consider all states that could be instantiated as $q'$ from Fig.~\ref{fig:super-linear} (lines~\ref{line:forQ}--\ref{line:Aa}).
 For each of these candidate $q'$'s, we  construct automata $\automaton_2, \automaton_3$ that correspond to paths $\pi_2, \pi_3$ from Fig.~\ref{fig:super-linear}  (lines~\ref{line:A2}--\ref{line:A3}).
Specifically, we construct  $\automaton_2$  by introducing a new initial state $q_i$
with transition  $(q_i, l, q_2)$ and making its accepting state $q'$.
Hence, $\automaton_2$ accepts strings that start in $q$, transition to $q_2$, and end in $q'$.

The construction of automaton $\automaton_3$, which should accept
all non-empty words that start and end in $q'$, 
is described in the {\sc AnyLoopBack} procedure. First, since we do not want $\automaton_3$ to accept
empty strings, we introduce a new initial state $q^\star$ and add a transition from $q^\star$ to all successor states $q_i$ of $q'$. Second, 
the final state of $\automaton'$ is $q'$ since we want to consider paths that loop back to $q'$.

The final missing piece of the algorithm is the construction of $\automaton_s$ (line~\ref{line:Aa}), whose complement accepts all attack suffixes for state
$q'$. As expected, $\automaton_s$ is the same as the original automaton $\automaton$, except that its initial state is $q'$. Finally, similar to Algorithm~\ref{Alg:algo1}, the attack automaton for states $q, q'$ is obtained as $\automaton_p \cdot
(\automaton_1 \cap \automaton_2 \cap \automaton_3)^+ \cdot
\overline{\automaton_s}$.
%
%
%
\begin{theorem}{\bf (Correctness of Algorithm~\ref{Alg:algo2})}
\label{thm:algo2}
  Let NFA $\automaton = (Q, \Sigma, \Delta, q_0, F)$ and  $\attack$ be the result of calling $\textsc{AttackAutomaton}(\automaton)$. 
  For every $s \in \langu(\attack)$, there exists a rejecting state $q_r \in Q \setminus F$ s.t. the
  number of distinct paths $\npath_i$ from $q_0$ to $q_r$ with
  $labels(\npath_i) = s$ is super-linear in the number of repetitions of the attack core in $s$.
\end{theorem}




\section{Dynamic Regular Expression Analysis}\label{sec:dynamic-regex}

Algorithms~\ref{Alg:algo1} and ~\ref{Alg:algo2} 
allow us to determine whether a given NFA is vulnerable. Even though our
static analyses are sound and complete at the NFA level, different regular
expression matching algorithms construct NFAs in different ways and use different backtracking search algorithms. Furthermore,
some matching algorithms may determinize the NFA (either lazily or eagerly) in
order to guarantee linear complexity. Since our analysis does not perform such
partial determinization of the NFA for a given regular expression, it can, in practice, generate
false positives. In addition, even if a regular expression is indeed
vulnerable, the input string must still exceed a certain minimum size to cause
denial-of-service.

In order to overcome these challenges in practice, we also perform dynamic
analysis to (a) confirm that a regular expression $\regex$ is indeed
vulnerable \emph{for Java's matching algorithm}, and (b) infer a minimum bound on the
size of the input string. Given the original regular expression $\regex$, a
user-provided time limit $t$, and the attack automaton $\attack$ (computed by
static regular expression analysis), our dynamic analysis produces a refined
attack automaton as well as a number $b$ such that there exists an input
string of length greater than $b$ for which Java's matching algorithm takes
more than $t$ seconds. Note that, as usual, this dynamic analysis trades soundness for completeness to avoid too many false
positives.

In more detail, given an attack automaton $\attack$ of the form $\automaton_p
\cdot \automaton_c^+ \cdot \automaton_s$, the dynamic analysis finds the
smallest $k$ where the shortest string $s \in \langu(\automaton_p \cdot
\automaton_c^k \cdot \automaton_s)$ exceeds the time limit $t$. In practice,
this process does not require more than a few iterations because we use the
complexity of the NFA to predict the number of repetitions that should be
necessary based on previous runs. The minimum required input length $b$ is
determined based on the length of the found string $s$. In addition, the value
$k$ is used to refine the attack automaton: in particular, given the original
attack automaton $\automaton_p \cdot \automaton_c^+ \cdot \automaton_s$, the
dynamic analysis refines it to be $\automaton_p \cdot \automaton_c^k \cdot
\automaton_c^* \cdot \automaton_s$.

\vspace{-5pt}

\section{Static Program Analysis}
\label{sec:informal_analysis}


As explained in Section~\ref{sec:overview}, the presence of a vulnerable regular expression does not necessarily mean that the program  is vulnerable.  In particular, there are three necessary conditions for the program to contain a ReDoS vulnerability:
First, a variable $x$ that stores user input  must be matched against a vulnerable regular expression $\regex$. Second,
it  must be possible for $x$ to store an attack string that triggers worst-case behavior for $\regex$; and, third,
the length of the string stored in $x$  must exceed the minimum threshold determined using dynamic analysis.



To determine if the program actually contains a ReDoS vulnerability, our approach also performs static analysis of source code.  Specifically, our program analysis employs  the Cartesian product~\cite{cartesian} of the following   abstract domains:


\begin{itemize}
\item The \emph{taint abstract domain}~\cite{flowdroid,TrippPistoia2009} tracks taint information for each  variable. In particular,  a variable is considered \emph{tainted} if it may store user input.
\item The \emph{automaton abstract domain}~\cite{YuAlkhalaf2014,YuAlkhalaf2010,ChristensenMoller2003} overapproximates the contents of string variables using finite automata. In particular, if string $s$ is in the language of automaton $\automaton$ representing  $x$'s contents, then $x$ \emph{may} store string~$s$.
\item The \emph{interval domain}~\cite{cousot77} is used to reason about string lengths. Specifically, we introduce a ghost variable $l_x$ representing the length of string  $x$ and use the interval abstract domain to infer upper and lower bounds for each $l_x$.

\end{itemize}

Since these abstract domains are fairly standard, we only explain how to use this information to detect ReDoS vulnerabilities. Consider a statement  ${\rm match}(x, \regex)$ that checks if string variable $x$ matches regular expression $\regex$, and suppose that the attack automaton for $\regex$ is  $\attack$. Now, our program analysis considers the statement  ${\rm match}(x, \regex)$ to be vulnerable if the following three conditions hold:

\begin{enumerate}
\item $\regex$ is vulnerable and variable $x$ is tainted;
\item The intersection of $\attack$ and the automaton abstraction of $x$ is non-empty;
\item The upper bound on ghost variable $l_x$ representing $x$'s length
  exceeds the minimum bound $b$ computed using dynamic analysis for $\attack$
  and a user-provided time limit $t$.
\end{enumerate}

\ifdefined\EXTENDEDVERSION
Appendix D offers a more rigorous formalization of the analysis.
\else
The extended version of this paper~\cite{WuestholzOlivo2017} offers a more rigorous formalization of the analysis.
\fi




\section{Experimental Evaluation}
\label{sec:expts}


To assess the usefulness of the techniques presented in this paper, we performed an evaluation in which our goal is to answer the following questions:
\vspace{-5pt}
\begin{itemize}
\item[\emph{\bf Q1:}] Do  real-world Java web applications  use vulnerable regular expressions?
\item[\emph{\bf Q2:}] Can \tool detect ReDoS vulnerabilities in web applications and how serious are these vulnerabilities?
\end{itemize}

\vspace{-7pt}


\vspace{0.1in} \noindent {\bf \emph{Results for Q1.}}  In order to assess
if real-world Java programs contain vulnerabilities, we scraped the top
$150$ Java web applications (by number of stars) that contain at least one
regular expression from GitHub repositories (all projects have between $10$ and
$2,000$ stars and at least $50$ commits) and collected a total of $2,864$ regular
expressions. In this pool of regular expressions, \tool found $37$ that have
worst-case exponential complexity and $522$ that have super-linear (but not
exponential) complexity.  Thus, we observe that approximately $20\%$ of the
regular expressions in the analyzed programs are vulnerable. We believe this
statistic highlights the need for more tools like \tool that can help
programmers reason about the complexity of regular expression matching.

\vspace{0.1in} \noindent
{\bf \emph{Results for Q2.}} To evaluate the effectiveness of \tool in finding ReDoS vulnerabilities, we
used \tool to statically analyze all Java applications that contain at least one vulnerable regular expression.
These programs include both web applications and frameworks,
 and cover a broad range of application domains. The average running time of \tool is approximately 14 minutes per program, including the time to dynamically analyze regular expressions. The average size of analyzed programs is about $58,000$ lines of code.

Our main result is that \tool found exploitable vulnerabilities in 27  applications
(including from popular projects, such as the Google Web Toolkit and Apache Wicket) and
reported a total of 46 warnings. We manually inspected each warning and confirmed that 41 out of the 46 vulnerabilities are exploitable, with 5 of the exploitable vulnerabilities involving hyper-vulnerable regular expressions and the rest being super-linear ones. \emph{Furthermore, for each of these 41 vulnerabilities (including super-linear ones), we were able to come up with a full, end-to-end exploit that causes the server to hang for more than 10 minutes.}



\begin{figure}[t]
\includegraphics[scale=0.18]{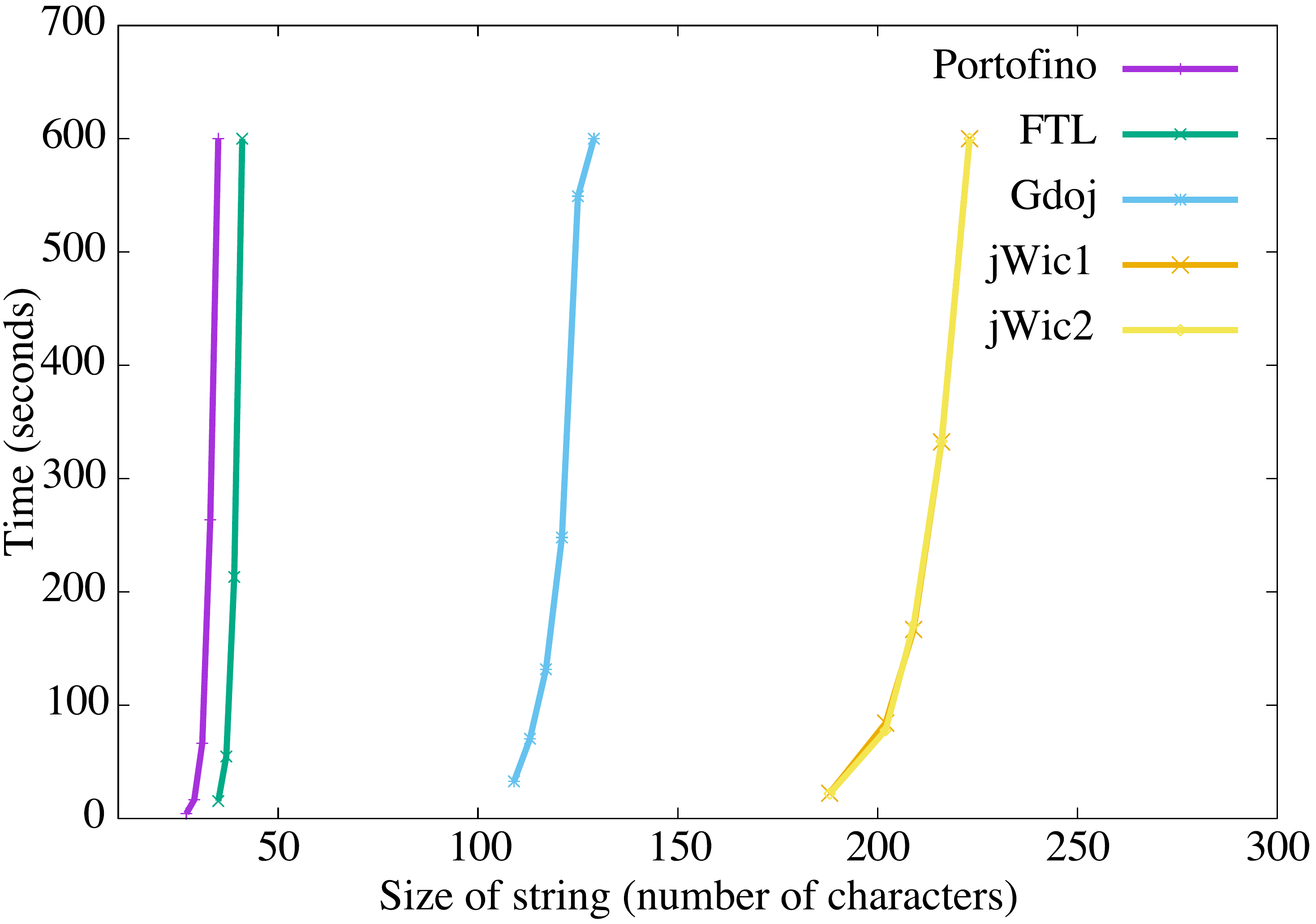}
\hfill
\includegraphics[scale=0.18]{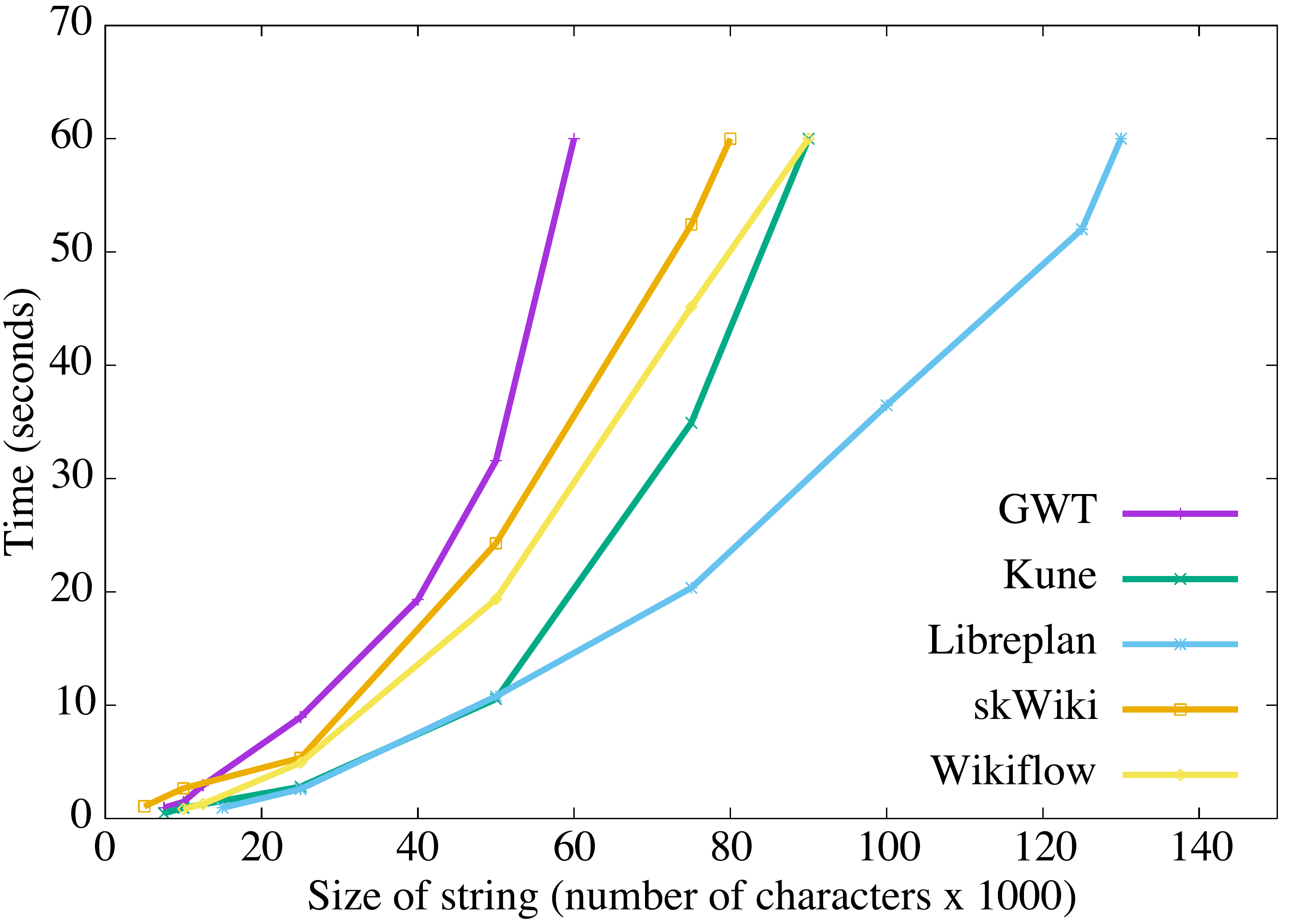}
\vspace{-0.1in}
\caption{Running times for exponential vulnerabilities (left) and super-linear vulnerabilities (right) for different input sizes.
}
\vspace{-0.2in}
\label{fig:eval}
\end{figure}

In Fig.~\ref{fig:eval}, we explore a subset of the vulnerabilities uncovered by \tool in more detail. Specifically,  Fig.~\ref{fig:eval} (left) plots input size against  running time for the {exponential vulnerabilities}, and
Fig.~\ref{fig:eval} (right) shows the same information  for a subset of the super-linear vulnerabilities.

\vspace{0.1in} \noindent
{\bf \emph{Possible fixes.}} We now briefly discuss some possible ways to fix the vulnerabilities uncovered by \tool. The most direct fix is to rewrite the regular expression so that it no longer exhibits super-linear complexity. Alternatively, the problem can also be fixed by ensuring that the user input cannot contain instances of the attack core. Since our technique provides the full attack automaton, we believe  \tool can be helpful for implementing suitable sanitizers. Another possible fix (which typically only works for super-linear regular expressions) is to bound input size. However, for most vulnerabilities found by \tool,  the input string can  legitimately be very large (e.g., review). Hence, there may not be an obvious upper bound,  or the bound may still be too large to  prevent a ReDoS  attack. For example, Amazon imposes an upper bound of 5000 words ($\sim$25,000 characters) on product reviews, but matching a super-linear regular expression against a string of that size may still take significant time. 



\section{Related Work}
\label{sec:related}


To the best of our knowledge, we are the first to present
an end-to-end solution for detecting ReDoS vulnerabilities by combining regular expression and program analysis. However,
there is prior work on static analysis of regular expressions and, separately, on program analysis  for finding security vulnerabilities. 


\vspace{0.1in} \noindent
{\bf \emph{Static analysis of regular expressions.}}
%
Since vulnerable regular expressions are known to be a significant problem, 
previous work has studied static analysis techniques for identifying 
regular expressions with worst-case exponential complexity~\cite{Berglund:2014,KirrageRathnayake2013,RathnayakeThielecke2014,Sugiyama:2014}.  
Recent work by Weideman et al.~\cite{Weideman:2016}
has also proposed an analysis for identifying super-linear regular expressions. However, no previous technique
can construct attack automata that capture all  malicious strings. Since attack automata are crucial for 
reasoning about sanitization, the algorithms we propose in this paper are necessary for performing sanitization-aware program analysis.
 Furthermore, we believe that the attack automata produced by our tool can help programmers write suitable sanitizers (especially in cases where the regular expression is difficult to rewrite).

\vspace{0.1in} \noindent
{\bf \emph{Program analysis for vulnerability detection.}}
%
There is a large body of work on statically detecting security vulnerabilities in programs.
Many of these techniques focus on detecting  cross-site scripting (XSS) or code  injection
vulnerabilities~\cite{BandhakaviTiku2011,ChaudhuriFoster2010,ChristensenMoller2003,DahseHolz2014,KiezunGuo2009,LivshitsLam2005,MartinLivshits2005,SuWassermann2006,Wassermann:2007,WassermannSu2008,WassermannYu2008,XieAiken2006,YuAlkhalaf2010,YuAlkhalaf2014,Yu:2011}. There has also been recent work on static detection of specific classes of denial-of-service vulnerabilities. For instance, Chang et al.~\cite{Chang:2009} and Huang et al.~\cite{Huang:2015} statically detect attacker-controlled loop bounds, and Olivo et al.~\cite{Olivo:2015} detect so-called \emph{second-order DoS vulnerabilities}, in which the size of a database query result is controlled by the attacker. However, as far as we know, there is no prior work that uses program analysis for detecting DoS vulnerabilities due to regular expression matching.

\vspace{0.1in} \noindent
{\bf \emph{Time-outs to prevent ReDoS.}} As mentioned earlier, some libraries (e.g., the {\sc .Net} framework) allow developers to set a time-limit for regular expression matching. While such libraries  may help \emph{mitigate} the problem through a band-aid solution, they do not address the root cause of the problem. For instance, they neither prevent against stack overflows nor do they prevent DoS attacks in which the attacker triggers the regular expression matcher many times.

\vspace{-7pt}

\section{Conclusions and Future Work}
\label{sec:conclusion}

We have presented an end-to-end solution for  statically detecting  regular expression denial-of-service vulnerabilities in programs. Our key idea is to combine complexity analysis of regular expressions with safety analysis of programs. Specifically, our regular expression analysis  constructs an attack automaton that recognizes all strings that trigger worst-case super-linear or exponential behavior. The program analysis component  takes this information as input and performs a combination of taint and string analysis to determine whether an attack string could be matched against a vulnerable regular expression.

We have used our tool to analyze thousands of regular expressions in the wild and we show that 20\% of regular expressions in the analyzed  programs are actually vulnerable. We also use \tool to analyze  Java web applications collected from Github repositories and find 41 exploitable security vulnerabilities in 27 applications.
 Each of these vulnerabilities  can be exploited to make the web server unresponsive for more than 10 minutes.

There are two main directions that we would like to explore in future work: First, we are interested in the problem of automatically \emph{repairing} vulnerable regular expressions. Since it is often difficult for humans to reason about the complexity of regular expression matching, we believe there is a real need for techniques that can automatically synthesize equivalent regular expressions with linear complexity. Second, we also plan to investigate the problem of automatically generating sanitizers from the attack automata produced by our regular expression analysis.

\bibliographystyle{splncs03}
\bibliography{evil-regexes}

\begin{thebibliography}{10}
\providecommand{\url}[1]{\texttt{#1}}
\providecommand{\urlprefix}{URL }

\bibitem{cve-superlinear1}
{CVE-2013-2009}. \url{cve.mitre.org/cgi-bin/cvename.cgi?name=CVE-2013-2099}

\bibitem{cve-net}
{CVE-2015-2525}. \url{cve.mitre.org/cgi-bin/cvename.cgi?name=CVE-2015-2526}

\bibitem{cve-exp3}
{CVE-2015-2525}. \url{cve.mitre.org/cgi-bin/cvename.cgi?name=CVE-2009-3275}

\bibitem{cve-superlinear2}
{CVE-2016-2515}. \url{cve.mitre.org/cgi-bin/cvename.cgi?name=CVE-2016-2515}

\bibitem{cve-superlinear3}
{CVE-2016-2537}. \url{cve.mitre.org/cgi-bin/cvename.cgi?name=CVE-2016-2537}

\bibitem{flowdroid}
Arzt, S., Rasthofer, S., Fritz, C., Bodden, E., Bartel, A., Klein, J., Traon,
  Y.L., Octeau, D., McDaniel, P.: Flowdroid: precise context, flow, field,
  object-sensitive and lifecycle-aware taint analysis for android apps. In:
  PLDI. pp. 259--269. {ACM} (2014)

\bibitem{cartesian}
Ball, T., Podelski, A., Rajamani, S.K.: Boolean and cartesian abstraction for
  model checking c programs. In: International Conference on Tools and
  Algorithms for the Construction and Analysis of Systems. pp. 268--283.
  Springer (2001)

\bibitem{BandhakaviTiku2011}
Bandhakavi, S., Tiku, N., Pittman, W., King, S.T., Madhusudan, P., Winslett,
  M.: Vetting browser extensions for security vulnerabilities with {VEX}.
  Commun. {ACM}  54(9),  91--99 (2011)

\bibitem{Berglund:2014}
Berglund, M., Drewes, F., van~der Merwe, B.: Analyzing catastrophic
  backtracking behavior in practical regular expression matching. In: AFL.
  {EPTCS}, vol. 151, pp. 109--123 (2014)

\bibitem{Chang:2009}
Chang, R.M., Jiang, G., Ivancic, F., Sankaranarayanan, S., Shmatikov, V.:
  Inputs of coma: Static detection of denial-of-service vulnerabilities. In:
  CSF. pp. 186--199. {IEEE} Computer Society (2009)

\bibitem{ChaudhuriFoster2010}
Chaudhuri, A., Foster, J.S.: Symbolic security analysis of ruby-on-rails web
  applications. In: CCS. pp. 585--594. {ACM} (2010)

\bibitem{ChristensenMoller2003}
Christensen, A.S., M{\o}ller, A., Schwartzbach, M.I.: Precise analysis of
  string expressions. In: SAS. LNCS, vol. 2694, pp. 1--18. Springer (2003)

\bibitem{cousot77}
Cousot, P., Cousot, R.: Abstract interpretation: a unified lattice model for
  static analysis of programs by construction or approximation of fixpoints.
  In: POPL. pp. 238--252. ACM (1977)

\bibitem{CrosbyWallach2003}
Crosby, S.A., Wallach, D.S.: Denial of service via algorithmic complexity
  attacks. In: {USENIX} Security Symposium. {USENIX} Association (2003)

\bibitem{DahseHolz2014}
Dahse, J., Holz, T.: Static detection of second-order vulnerabilities in web
  applications. In: {USENIX} Security Symposium. pp. 989--1003. {USENIX}
  Association (2014)

\bibitem{Huang:2015}
Huang, H., Zhu, S., Chen, K., Liu, P.: From system services freezing to system
  server shutdown in android: All you need is a loop in an app. In: CCS. pp.
  1236--1247. ACM (2015)

\bibitem{KiezunGuo2009}
Kiezun, A., Guo, P.J., Jayaraman, K., Ernst, M.D.: Automatic creation of {SQL}
  injection and cross-site scripting attacks. In: ICSE. pp. 199--209. {IEEE}
  (2009)

\bibitem{KirrageRathnayake2013}
Kirrage, J., Rathnayake, A., Thielecke, H.: Static analysis for regular
  expression denial-of-service attacks. In: NSS. LNCS, vol. 7873, pp. 135--148.
  Springer (2013)

\bibitem{LivshitsLam2005}
Livshits, V.B., Lam, M.S.: Finding security vulnerabilities in java
  applications with static analysis. In: {USENIX} Security Symposium. {USENIX}
  Association (2005)

\bibitem{MartinLivshits2005}
Martin, M.C., Livshits, V.B., Lam, M.S.: Finding application errors and
  security flaws using {PQL:} a program query language. In: OOPSLA. pp.
  365--383. {ACM} (2005)

\bibitem{Olivo:2015}
Olivo, O., Dillig, I., Lin, C.: Detecting and exploiting second order
  denial-of-service vulnerabilities in web applications. In: CCS. pp. 616--628.
  {ACM} (2015)

\bibitem{RathnayakeThielecke2014}
Rathnayake, A., Thielecke, H.: Static analysis for regular expression
  exponential runtime via substructural logics. CoRR  abs/1405.7058 (2014)

\bibitem{SuWassermann2006}
Su, Z., Wassermann, G.: The essence of command injection attacks in web
  applications. In: POPL. pp. 372--382. {ACM} (2006)

\bibitem{Sugiyama:2014}
Sugiyama, S., Minamide, Y.: Checking time linearity of regular expression
  matching based on backtracking. IPSJ Online Transactions  7,  82--92 (2014)

\bibitem{thompson}
Thompson, K.: Programming techniques: Regular expression search algorithm.
  Communications of the ACM  11(6),  419--422 (1968)

\bibitem{TrippPistoia2009}
Tripp, O., Pistoia, M., Fink, S.J., Sridharan, M., Weisman, O.: {TAJ:}
  effective taint analysis of web applications. In: PLDI. pp. 87--97. {ACM}
  (2009)

\bibitem{Wassermann:2007}
Wassermann, G., Su, Z.: Sound and precise analysis of web applications for
  injection vulnerabilities. In: PLDI. pp. 32--41. {ACM} (2007)

\bibitem{WassermannSu2008}
Wassermann, G., Su, Z.: Static detection of cross-site scripting
  vulnerabilities. In: ICSE. pp. 171--180. {ACM} (2008)

\bibitem{WassermannYu2008}
Wassermann, G., Yu, D., Chander, A., Dhurjati, D., Inamura, H., Su, Z.: Dynamic
  test input generation for web applications. In: ISSTA. pp. 249--260. {ACM}
  (2008)

\bibitem{Weideman:2016}
Weideman, N., van Der~Merwe, B., Berglund, M., Watson, B.: Analyzing matching
  time behavior of backtracking regular expression matchers by using ambiguity
  of {NFA}. In: CIAA (2016), to appear

\bibitem{XieAiken2006}
Xie, Y., Aiken, A.: Static detection of security vulnerabilities in scripting
  languages. In: {USENIX} Security Symposium. {USENIX} Association (2006)

\bibitem{YuAlkhalaf2010}
Yu, F., Alkhalaf, M., Bultan, T.: Stranger: An automata-based string analysis
  tool for {PHP}. In: TACAS. LNCS, vol. 6015, pp. 154--157. Springer (2010)

\bibitem{YuAlkhalaf2014}
Yu, F., Alkhalaf, M., Bultan, T., Ibarra, O.H.: Automata-based symbolic string
  analysis for vulnerability detection. FMSD  44(1),  44--70 (2014)

\bibitem{Yu:2011}
Yu, F., Bultan, T., Hardekopf, B.: String abstractions for string verification.
  In: SPIN. LNCS, vol. 6823, pp. 20--37. Springer (2011)

\end{thebibliography}

\ifdefined\EXTENDEDVERSION
\appendix
\section*{Appendix A: Proof about Algorithm~\ref{Alg:algo1}}
\label{sec:proof_algo1}

\begin{proof} (sketch)
 We first show that any attack string $s$ is accepted by  $\attack$. Based on Section~\ref{sec:regex}, we know that attack strings that cause exponential behavior are of the form $s_0 \cdot s_c^k \cdot s_1$ where $s_0 = \labels(\pi_p)$, $s_c = \labels(\pi_1) = \labels(\pi_2)$, $s_1 = \labels(\pi_s)$ for some pivot state $q$.
  Now, we argue that $s$ will be accepted by the attack automaton
  $\attack_q$ for $q$, which implies that $s$ is also accepted by $\attack$ since
  $\attack_q \subseteq \attack$. Since \textsc{AttackForPivot} is invoked for each state $q$,
  we will consider the two distinct transitions $(q, l, q_1)$ and $(q, l, q_2)$ that start paths
  $\npath_1$ and $\npath_2$.  Furthermore, by the construction in the {\sc LoopBack} procedure,
  $\labels(\npath_1)$ and $\labels(\npath_2)$ will be accepted by $\automaton_1$ and $\automaton_2$. Thus, string $s_c$ will be accepted by $(\automaton_1 \cap \automaton_2)$. Similarly, by the construction at lines 14--15, $\automaton_p$ and $\overline{\automaton_s}$ will accept $s_0$ and $s_1$ respectively. Hence, the attack string $s = s_0 s_c^k s_1$ will be recognized by
  $\automaton_p \cdot (\automaton_1 \cap \automaton_2)^+ \cdot
  \overline{\automaton_s}$.

  For the other direction, we show that if a string $s$ is accepted by
  $\attack$, then it is indeed an attack string. The attack automaton
  constructed by the algorithm is a union of automata $\attack_q$ of the form
  $\automaton_p \cdot (\automaton_1 \cap \automaton_2)^+ \cdot
  \overline{\automaton_s}$, one of which must accept the string $s$. As a
  consequence, there must exist strings $s_0 \in \automaton_p$, $s_c^k \in
  (\automaton_1 \cap \automaton_2)^+$, and $s_1 \in \overline{\automaton_s}$
  such that $s = s_0 \cdot s_c^k \cdot s_1$. Due to the construction of
  $\attack_q$ in function \textsc{AttackForPivot}, there must exist
  corresponding paths $\pi_1$, $\pi_2$ (distinct from $\pi_1$), $\pi_p$, and
  $\pi_s$ such that $s_0 = \labels(\pi_p)$, $s_c = \labels(\pi_1)
  = \labels(\pi_2)$, and $s_1 = \labels(\pi_s)$.  Based on
  Section~\ref{sec:regex}, any such string $s$ constitutes an attack
  string.
\end{proof}

\section*{Appendix B: Proof about Algorithm~\ref{Alg:algo2}}
\label{sec:proof_algo2}

\begin{proof} (sketch)
  We first show that any attack string $s$ is accepted by $\attack$. Based on
  Section~\ref{subsec:vulnerable-nfa-super-linear}, we know that attack strings that cause
  super-linear behavior are of the form $s_0 \cdot s_c^k \cdot s_1$ where $s_0
  = \labels(\pi_p)$, $s_c = \labels(\pi_1) = \labels(\pi_2) = \labels(\pi_3)$,
  $s_1 = \labels(\pi_s)$ for some pivot state $q$ and a state $q'$. Now, we
  argue that $s$ will be accepted by the attack automaton $\attack_q$ for $q$,
  which implies that $s$ is also accepted by $\attack$ since $\attack_q
  \subseteq \attack$. Since \textsc{AttackForPivot} is invoked for each state
  $q$, we will consider the two distinct transitions $(q, l, q_1)$ and $(q, l,
  q_2)$ that start paths $\npath_1$ and $\npath_2$ and any state
  $q'$. Furthermore, by the construction in the {\sc LoopBack} procedure,
  $\labels(\npath_1)$ will be accepted by $\automaton_1$. By the construction
  on lines~\ref{line:forQ}--\ref{line:A2}, $\labels(\npath_2)$ will be
  accepted by $\automaton_2$. By the construction in the \textsc{AnyLoopBack}
  procedure, $\labels(\npath_3)$ will be accepted by $\automaton_3$.  Thus,
  string $s_c$ will be accepted by $(\automaton_1 \cap \automaton_2 \cap
  \automaton_3)$. Similarly, by the construction at lines~\ref{line:Ap}
  and~\ref{line:As}, $\automaton_p$ and $\overline{\automaton_s}$ will accept
  $s_0$ and $s_1$ respectively. Hence, the attack string $s = s_0 s_c^k s_1$
  will be recognized by $\automaton_p \cdot (\automaton_1 \cap \automaton_2
  \cap \automaton_3)^+ \cdot \overline{\automaton_s}$.

  For the other direction, we show that if a string $s$ is accepted by
  $\attack$, then it is indeed an attack string. The attack automaton
  constructed by the algorithm is a union of automata $\attack_q$ of the form
  $\automaton_p \cdot (\automaton_1 \cap \automaton_2 \cap \automaton_3)^+
  \cdot \overline{\automaton_s}$, one of which must accept the string $s$. As
  a consequence, there must exist strings $s_0 \in \automaton_p$, $s_c^k \in
  (\automaton_1 \cap \automaton_2 \cap \automaton_3)^+$, and $s_1 \in
  \overline{\automaton_s}$ such that $s = s_0 \cdot s_c^k \cdot s_1$. Due to
  the construction of $\attack_q$ in function \textsc{AttackForPivot}, there
  must exist corresponding paths $\pi_1$, $\pi_2$ (distinct from $\pi_1$),
  $\pi_3$, $\pi_p$, and $\pi_s$ such that $s_0 = \labels(\pi_p)$, $s_c
  = \labels(\pi_1) = \labels(\pi_2) = \labels(\pi_3)$, and $s_1
  = \labels(\pi_s)$.  Based on Section~\ref{subsec:vulnerable-nfa-super-linear} any such
  string $s$ constitutes an attack string.
\end{proof}


\section*{Appendix C: Necessity Proofs (Theorems~\ref{thm:hyper-vulnerable} and~\ref{thm:vulnerable-super-linear})}
\label{sec:complexity}

This section shows that the conditions in Theorems~\ref{thm:hyper-vulnerable} and~\ref{thm:vulnerable-super-linear} 
are not only sufficient, but also necessary for the NFA to exhibit exponential and super-linear complexity respectively.  In the rest of this section, we use the term \emph{vulnerable NFA} to mean an NFA that satisfies the conditions of Theorem~\ref{thm:hyper-vulnerable} and \emph{hyper-vulnerable NFA} to mean an NFA satisfying conditions of Theorem~\ref{thm:vulnerable-super-linear}.

Our proof uses the concept of {\em strongly-connected component 
pair path} (SPP). Given an NFA with $|Q|$ states, there are at most $\mathcal{O}(|Q|^{|Q|})$ such SPPs. 
Given a string $s$, if the NFA is not hyper-vulnerable, then there are at most $|s|^{|Q|}$ possible matchings per
SPP. The complexity of a worst-case backtracking search algorithm is thus polynomial in length of $s$: $\mathcal{O}(|Q|^{|Q|}|s|^{|Q|})$.

If the NFA is not vulnerable, then there are at most $|Q|^{|Q|}$ possible matchings per SPP.
A backtracking search algorithm could match all possible substrings of $s$ adding a of factor $|s|$ and 
a factor of $|\Sigma|^{|Q|}$, resulting in the complexity: $\mathcal{O}(|Q|^{|Q|}|\Sigma|^{|Q|}|s|)$.


\begin{definition}{\bf (Strongly-Connected Component)}
\label{def:scc}
Given an NFA\\$\automaton = (Q, \Sigma, \Delta, q_0, F)$. Two states $q, q' \in Q$ with $q \neq q'$ are in the 
same strongly-connected component if and only if there exist a path from $q$ to $q'$ and from $q'$ to $q$.
A partition of $Q$ in strongly connected components is unique. 
\end{definition}

\begin{lemma}
\label{lem:unique}
Given a non-hyper-vulnerable NFA $\automaton = (Q, \Sigma, \Delta, q_0, F)$, a string $s$, and two states $q, q' \in Q$ 
occurring in the same strongly-connected component. The path $\npath$ from $q$ to $q'$ such that $\labels(\npath) = s$
is unique.
\end{lemma}

\begin{proof}
Assume that there are two paths $\npath_1$ and $\npath_2$ from $q$ to $q'$ such that $\npath_1 \neq \npath_2$
and $\labels(\npath_1) = \labels(\npath_2) = s$. Since $q$ and $q'$ occur in the same strongly-connected component, 
there must be a path $\npath_3$ from $q'$ to $q$, because in a strongly-connected component there exists a path from
every state to every state. Now we have two cycles from $q$ to $q$, $\npath_1\npath_3$ and $\npath_2\npath_3$ such
that $\labels(\npath_1\npath_3) = \labels(\npath_2\npath_3)$. This violates the assumption that $\automaton$ is 
not hyper-vulnerable. 
\end{proof}

Notice that Lemma~\ref{lem:unique} also holds for non-vulnerable NFAs as each non-vulnerable NFA is also 
non-hyper-vulnerable.

\begin{definition}{\bf (Strongly-Connected Component Pair Path)}
\label{def:scc}
Given an NFA $\automaton = (Q, \Sigma, \Delta, q_0, F)$. A strongly-connected component pair path (SPP)
of $\automaton$ is a sequence of state pairs $(q_{\mathrm{in}}, q_{\mathrm{out}})$, with $q_{\mathrm{in}}$
and $q_{\mathrm{out}}$ occurring in the same strongly-connected component. Moreover, for any two 
consecutive (i.e., occurring in different, but connected strongly-connected components) state pairs
$(q_{\mathrm{in}}, q_{\mathrm{out}})$ and $(q'_{\mathrm{in}}, q'_{\mathrm{out}})$ there
must exist a transition $(q_{\mathrm{out}}, l, q'_{\mathrm{in}}) \in \Delta$ for some label $l$.
\end{definition}

\begin{lemma}
\label{lem:spp}
Given an NFA $\automaton = (Q, \Sigma, \Delta, q_0, F)$. There are at most $\mathcal{O}(|Q|^{|Q|})$
different strongly-connected component pair paths.
\end{lemma}

\begin{proof}
$\automaton$ has at most $|Q|$ strongly-connected components. Consequently, a SPP
consists of at most $|Q|$ pairs. Each strongly-connected component consists of at most 
$|Q|$ states. Hence there are at most $|Q|^2$ different state pairs per strongly-connected
components. The number of SPPs for $\automaton$ is thus at most $|Q|^{2|Q|}$ or
$\mathcal{O}(|Q|^{|Q|})$.
\end{proof}

\begin{definition}{\bf (Path Partition)}
\label{def:pp}
Given an NFA $\automaton = (Q, \Sigma, \Delta, q_0, F)$, a string $s$ and
a strongly-connected component pair path $\Pi$ of $\automaton$.
A {\em path partition} of $s$ and $\Pi$ is a partition of $s$ into $2|\Pi| - 1$ substrings 
$s_i$ with $i \in \{0, \dots, 2|\Pi| - 2\}$ in such a way that each $s_i$ with odd $i$ consists of exactly one symbol. 
The substrings $s_i$ with $i$ even can be arbitrarily long (or short, even empty). Notice that the choice of the 
$s_i$ with odd $i$ define the $s_i$ with the even $i$: Let $i$ be odd, $s_{i+1}$ are all
the symbols in $s$ that occur between $s_i$ and $s_{i+2}$.
\end{definition}

\begin{lemma}
\label{lem:pp}
Given an NFA $\automaton = (Q, \Sigma, \Delta, q_0, F)$, a string $s$ and
a strongly-connected component pair path $\Pi$ of $\automaton$. There exist
at most $|s|^{|Q|}$ different path partitions of $s$ and $\Pi$.
\end{lemma}

\begin{proof}
There exists at most $\binom {|s|} {|\Pi| - 1}$ different path partitions of
 $s$ and $\Pi$, i.e., all possible $|\Pi| - 1$ choices of $s_i$'s with odd $i$. Notice that 
 $|\Pi| - 1 < |Q|$, because a SPP has at most length $|Q|$. Therefore the number 
 of different path partitions is less than $|s|^{|Q|}$.
\end{proof}

\begin{theorem}
  \label{thm:poly-in-s}
  Let $\automaton$ be the NFA $(Q, \Sigma, \Delta, q_0, F)$, which is not hyper-vulnerable. 
  The runtime to determine if a string $s$ is accepted by $\automaton$ is
  at most $\mathcal{O}(|Q|^{|Q|}|s|^{|Q|})$. 
\end{theorem}

\begin{proof}
Below we assume that $s$ consists of at least $|Q|$ symbols. In case $|s| < |Q|$, 
then the number of steps is limited by $|Q|^{|s|}$ and thus $|Q|^{|Q|}$ even for hyper-vulnerable NFAs: 
From each state we could potentially go to each other state and repeat that $|s|$ times.


From Lemma~\ref{lem:unique} we know that there is a path from state $q$ to $q'$ with 
$q$ and $q'$ occurring in the same strongly connected component is unique. Given 
path partition $P$ of $s$ and $\Pi$, we can now deduce that the path of $s$ from $q_0$
to the last $q_{\mathrm{out}}$ in $\Pi$ is unique: each $s_i$ with even $i$ uniquely 
force the path within a strongly-connected component, while each $s_i$ with odd $i$ 
uniquely define the path in between strongly-connected components. 

Furthermore, from Lemma~\ref{lem:pp}, we know that the number of part partitions is
at most $|s|^{|Q|}$.

Consequently, a backtracking search algorithm for $s$ will require at most 
$\mathcal{O}(|Q|^{|Q|}|s|^{|Q|})$ steps. Notice that we did not discuss that a backtrack
search algorithm also matches strings that are the first $n$ symbols of $s$. This
adds another factor of $|s|$, which can be ignored given the above complexity result.
\end{proof}





\begin{definition}{\bf (Labelled Strongly-Connected Component Pair Path)}
\label{def:lspp}
Given an NFA $\automaton = (Q, \Sigma, \Delta, q_0, F)$. A labelled strongly-connected component pair path (LSPP)
of $\automaton$ is a strongly-connected component pair path with has a specific label $l$ in between two
consecutive state pairs $(q_{\mathrm{in}}, q_{\mathrm{out}})$ and $(q'_{\mathrm{in}}, q'_{\mathrm{out}})$ that 
describes the transition from $q_{\mathrm{out}}$ to $q'_{\mathrm{in}}$.
\end{definition}

\begin{lemma}
\label{lem:lspp}
Given an NFA $\automaton = (Q, \Sigma, \Delta, q_0, F)$. There are at most $\mathcal{O}(|Q|^{|Q|}|\Sigma|^{|Q|})$
different labelled strongly-connected component pair paths.
\end{lemma}

\begin{proof}
Let $\Pi$ be a strongly-connected component pair path. There are at most $\mathcal{O}(|Q|^{|Q|})$ 
strongly-connected component pair paths (Lemma~\ref{lem:spp}). There are $|\Pi| - 1$ consecutive state 
pairs in $\Pi$. For each of them a label $l \in \Sigma$ can be selected. Hence there are 
$|\Sigma|^{|\Pi| - 1}$ different labelled strongly-connected component pair path that have the 
same state pairs as $\Pi$. This results in $\mathcal{O}(|Q|^{|Q|}|\Sigma|^{|Q|})$ different LSSPs.
\end{proof}

A path partition $P$ of string $s$ and strongly-connected component pair path $\Pi$ 
is called {\em valid}, if there exists a path $\npath$ that follows the states described 
in $\Pi$ such that $\labels(\npath) = s$. Lemma~\ref{lem:pp} states that for a 
non-hyper-vulnerable NFA  there are at most $|s|^{|Q|}$ path partitions. All of them 
could be valid. However, below we will show that for a non-vulnerable NFA that the 
number of valid path partitions is at most $|Q|^{|Q|}$, thereby removing the factor
$|s|^{|Q|}$ from the complexity.

\begin{lemma}
\label{lem:non-vulnerable}
Given a non-vulnerable NFA $\automaton = (Q, \Sigma, \Delta, q_0, F)$, a string $s$,
and a labelled strongly-connected component pair path $\Pi$. 
Let $(q_{in}, q_{\mathrm{out}})$ and $(q'_{\mathrm{in}}, q'_{\mathrm{out}})$ 
be two consecutive state pairs of $\Pi$ and let $l$ be the label in between the state pairs:
$(q_{\mathrm{out}}, l, q'_{\mathrm{in}}) \in \Delta$.
There are at most $|Q|$ possible choices of $l$ in $s$, such that there exists
a valid path partition $P$ of $s$ and $\Pi$.
\end{lemma}

\begin{proof} (sketch)
Let $q_{\mathrm{end}}$ be the last state in $\Pi$, i.e., the second state in the last state pair. 
Let $Q_0$ be all the states for which a path exists to $q_{\mathrm{out}}$ including $q_{\mathrm{out}}$ and 
let $Q_1$ be all states for which a path exists starting from $q'_{\mathrm{in}}$ including $q'_{\mathrm{in}}$.
Notice that the intersection of $Q_0$ and $Q_1$ is empty.

Consider $|Q_0|$ different choices of $l$ in $s$ such that exists a path from 
$q_0$ to $q_{\mathrm{out}}$ following the states described in $\Pi$ and a 
path from $q'_{\mathrm{in}}$ to $q_{\mathrm{end}}$ following the states described in $\Pi$.
On any such path that can be at most $|Q_0| - 1$ different $l$ transitions. Hence there 
must exists a chosen transition $(q, l, q')$ such that  the path from $q_0$ to $q_{\mathrm{out}}$ uses that 
transition at least twice. For this $q$, there exists a cycle $\pi_1$ from $q$ to $q$ and a path $\pi_2$ starting at $q$ 
that includes the transition $(q_{\mathrm{out}}, l, q'_{\mathrm{in}})$ such that $\labels(\pi_1) = \labels(\pi_2)$.

We can apply the same reason for $Q_1$: Consider $|Q_1|$ different picks of $l$ in $s$ such that exists 
a path from $q_0$ to $q_{\mathrm{out}}$ following the states described in $\Pi$ and a 
path from $q'_{\mathrm{in}}$ to $q_{\mathrm{end}}$ following the states described in $\Pi$.
On any such path that can be at most $|Q_1| - 1$ different $l$ transitions. Hence there 
must exists a transition $(q'', l, q''')$ such that the path from $q'_{\mathrm{in}}$ to $q_{\mathrm{end}}$ uses that 
transition at least twice. For this $q''$, there exists a cycle $\pi_3$ from $q''$ to $q''$ and a path $\pi_4$ ending 
at $q''$ that includes the transition $(q_{\mathrm{out}}, l, q'_{\mathrm{in}})$ such that $\labels(\pi_3) = \labels(\pi_4)$.

Let $\pi_5$ be a path from $q$ to $q''$. Now we can change the cycles $\pi_1$ and $\pi_3$ to 
$\pi'_1$ and $\pi'_3$ by extending them with loops such that $\labels(\pi'_1) = \labels(\pi'_3) = \labels(\pi_5)$.
The existence of such paths is in conflict with the assumption that $\automaton$ is not vulnerable.
Consequently, there must be less than $|Q_0| + |Q_1|$ choices for $l$. Since 
the intersection of $Q_0$ and $Q_1$ is empty, $|Q_0| + |Q_1| \leq |Q|$.
\end{proof}

\begin{theorem}
  \label{thm:linear-in-s}
  Let $\automaton$ be the NFA $(Q, \Sigma, \Delta, q_0, F)$, which is not vulnerable. 
  The runtime to determine if a string $s$ is accepted by $\automaton$ is
  at most $\mathcal{O}(|Q|^{|Q|}|\Sigma|^{|Q|}|s|)$.
\end{theorem}

\begin{proof}
From Lemma~\ref{lem:lspp} we know that there are at most $\mathcal{O}(|Q|^{|Q|}|\Sigma|^{|Q|})$
labelled strongly-connected component pair path of $s$. Let $\Pi$ be one of these LSPPs.
There are $|\Pi| - 1$ labels between consecutive state pairs. For each of them there are at 
lost $|Q|$ choices from $s$, such that the path partition is valid (Lemma~\ref{lem:non-vulnerable}). 
Consequently, there are at most $|Q|^{|\Pi| - 1}$ valid path partitions for $\Pi$. Since $|\Pi| - 1 < |Q|$, 
the number of valid path 
partitions is less than $|Q|^{|Q|}$. Hence we can ignore the number of valid path partitions 
in the complexity, because it does not alter $\mathcal{O}(|Q|^{|Q|}|\Sigma|^{|Q|})$. 

Given a valid path partition, a backtrack search algorithm may take $|s|$ steps from $q_0$ to
the last state. This adds a factor of $|s|$ to the complexity resulting in $\mathcal{O}(|Q|^{|Q|}|\Sigma|^{|Q|}|s|)$.

\end{proof}

\newcommand{\fst}[1]{\emph{fst}(#1)}
\newcommand{\snd}[1]{\emph{snd}(#1)}

\section*{Appendix D: Formal Analysis}
\label{sec:analysis}

\begin{figure}
\[
\begin{array}{lll}
{\rm Statement} \ S & := & \ x := e \ | \ {\rm getInput}(x) \\
& & | \ {\rm match}(x, \regex) \ | \ S_1; S_2 \\
& & | \ {\rm assume}(x \in \impure) \\
& & | \ {\rm assume}(\len{x} \leq \nu)  \\
& & | \ {\rm if}(\star) \ {\rm then} \ S_1 \ {\rm else} \ S_2 \\
& & | \ {\rm while}(\star) \ {\rm do} \ S \\
& \\
{\rm String \ exp} \ e & := & x \ | \ ? \\
{\rm Int \ exp} \  \nu & := & {\rm int} \ | \ \len{x} \ | \ \nu_1 + \nu_2 \ | \ \nu_1 - \nu_2  \\  
{\rm Pure \ regex} \ \regex & := & a \in \Sigma \ | \  \regex^* \ | \ \regex_1 + \regex_2 \ | \ \regex_1 \cdot \regex_2 \\
{\rm Impure \ regex} \ \impure & := & \regex \ | \ x \ |  \  \impure^* \ | \ \impure_1 + \impure_2 \ | \ \impure_1 \cdot \impure_2 \\
\end{array}
\]
\vspace{-0.1in}
\caption{The \il intermediate language }
\vspace{-0.1in}
\label{fig:lang}
\end{figure}

\noindent
{\bf \emph{Intermediate language.}} We formalize our program analysis using the intermediate language shown in Fig.~\ref{fig:lang}. 
This language, which we refer to as \il, is suitable for describing our analysis because it  models the effects of different string manipulation functions in a uniform way using \emph{assume} statements.

In \il, all variables have type string. In addition to the standard  assignment, sequence, conditional, and loop constructs,  \il contains a  function  \emph{getInput(x)}, which binds variable $x$ to a string supplied by the user. Another  function, \emph{match($x, \regex$)},  models matching string $x$ against regular expression~$\regex$.  

The \il language contains two kinds of \emph{assume} statements that allow us to model the effect of string manipulation procedures (e.g., provided by {\tt java.lang.String}). First, the statement $\emph{assume}(x \in \impure)$ states that $x$ belongs to the language given by \emph{impure regular expression} $\impure$. Here, we refer to $\impure$ as \emph{impure} because the regular expression can refer to program variables. For example, the statement $\emph{assume}(x \in y \cdot a)$ models that the value stored in $x$ is the value stored in $y$ concatanated with the character $a$. Thus, if $y$ can be any string, then this annotation expresses that $x$ is a string ending in $a$. The use of such impure regular expressions in \il allows us to  model string operations in a uniform way.

The second form of annotation in \il is of the form $\emph{assume}(\emph{len}(x) \leq \nu)$ and allows us to express constraints on the size of strings. Here, the integer expression $\nu$ can refer to the length of other strings and can contain arithmetic operators ($+, -$).

One final point worth noting is that string expressions include a special symbol $?$, which represents an unknown string constant. Hence, an assignment of the form {\tt x = "abc"} is easily expressible in our language using the code snippet:
\[ x := ?;  \ {\rm assume}(x \in abc) \]

\begin{figure}[t]
\[
\begin{array}{rll}
\evalregex{\regex}{\Lambda} & = & \automaton(\regex)  \\
\evalregex{x}{\Lambda} & = & \emph{snd}(\Lambda(x)) \\
\evalregex{\impure^*}{\Lambda} & = & (\evalregex{\impure}{\Lambda})^* \\
\evalregex{\impure_1 \impure_2}{\Lambda} & = & \evalregex{\impure_1}{\Lambda} \cdot \evalregex{\impure_2}{\Lambda} \\
\evalregex{\impure_1 + \impure_2}{\Lambda} & = & \evalregex{\impure_1}{\Lambda} + \evalregex{\impure_2}{\Lambda} \\
\end{array}
\]
\vspace{-0.1in}
\caption{Helper rules for evaluating impure regular expressions. We use $\automaton(\regex)$ to denote an NFA that accepts the same language as regular expression $\regex$.}\label{fig:eval-impure}
\vspace{-0.1in}
\end{figure}

\begin{figure}[t]
\[
\begin{array}{rll}
\evalint{{\rm int}}{\Lambda} & = & \interval{{\rm int}}{{\rm int}}  \\
\evalint{\len{x}}{\Lambda} & = & \emph{fst}(\Lambda(x)) \\
\evalint{\nu_1 + \nu_2}{\Lambda} & = & \evalint{\nu_1}{\Lambda} \oplus  \evalint{\nu_2}{\Lambda} \\
\evalint{\nu_1 - \nu_2}{\Lambda} & = & \evalint{\nu_1}{\Lambda} \ominus  \evalint{\nu_2}{\Lambda} \\
\end{array}
\]
\vspace{-0.1in}
\caption{Helper rules for evaluating arithmetic expressions}\label{fig:eval-int}
\vspace{-0.1in}
\end{figure}

\paragraph{Program abstraction.} As mentioned earlier, our program analysis needs to track taint information as well as information about string lengths and contents. Towards this goal, our analysis employs three kinds of program abstractions:
\begin{itemize}
\item The \emph{taint abstraction} $\Phi$ is a set of variables such that $x \in \Phi$ indicates that $x$ may be tainted.
\item The \emph{string abstraction} $\Lambda$ is a mapping from each program variable $x$ to a pair $(\interv, \automaton)$, where $\interv$ is an interval $\interval{l}{u}$ such that $l \leq \emph{len}(s) \leq u$ and $\automaton$ is an NFA representing $x$'s contents. In particular, if a string $s$ is in the language of $\automaton$, this indicates that $x$ can store string $s$.
\item The \emph{attack abstraction} $\Psi$ maps each regular expression $\regex$ in the program to a pair $(b, \attack)$. Here, $b$ is a minimum bound on the  length of the input string $s$ such that, if $\emph{len}(s) < b$, matching $s$ against $\regex$ takes negligible time~\footnote{Here, what constitutes \emph{negligible time} is an input parameter of our analysis and can be customized by the user.}. The NFA $\attack$ is the attack automaton for regular expression $\regex$ and is pre-computed using the 
analyses from Sections~\ref{sec:regex} and ~\ref{sec:regex-superlinear}.
\end{itemize}

Since our program abstractions involve pairs (e.g., $(\interv, \automaton)$) , we use the notation \emph{fst(p)} and \emph{snd(p)} to retrieve the first and second components of pair $p$ respectively.

\begin{figure}[t]
\[
\begin{array}{rll}
\interval{l_1}{u_1} \oplus \interval{l_2}{u_2} & = &  \interval{l_1+l_2}{u_1+u_2}  \\
\interval{l_1}{u_1} \ominus \interval{l_2}{u_2} & = &  \interval{l_1-l_2}{u_1-u_2}  \\
\interval{l_1}{u_1} \sqcup \interval{l_2}{u_2} & = &  \interval{\emph{min}(l_1, l_2)}{\emph{max}(u_1, u_2)}  \\
(\Lambda_1 \sqcup \Lambda_2)(x) & = & (\fst{\Lambda_1(x)} \sqcup \fst{\Lambda_2(x)}, \\
& & \ \snd{\Lambda_1(x)} \cup \snd{\Lambda_2(x)})
\end{array}
\]
\vspace{-0.1in}
\caption{Operations on abstract domains}\label{fig:interval-ops}
\vspace{-0.1in}
\end{figure}

\paragraph{Analysis rules.} We describe our static analysis using judgments of the form
$
\Psi, \Phi, \Lambda \vdash S: \Phi', \Lambda'
$
which state that, if we execute $S$ in a state that satisfies program abstractions $\Psi, \Phi, \Lambda$, we obtain a new taint abstraction $\Phi'$ and new string abstraction $\Lambda'$. The inference rules describing our analysis are shown in Fig.~\ref{fig:rules}.

\begin{figure}
\[
\begin{array}{cc}
(1) & \irule{
\begin{array}{c}
\Phi' = \Phi \cup \{ x \} \\
\Lambda' = \Lambda[x \mapsto (\interval{0}{\infty}, \automaton^*)]
\end{array}
}{
\Psi, \Phi, \Lambda \vdash {\rm getInput}(x): \Phi', \Lambda'
} \\ \ \\
(2) & \irule{
\begin{array}{c}
\Lambda' = \Lambda[x_1 \mapsto \Lambda(x_2)] \\
\Phi' = \left \{
\begin{array}{ll}
\Phi \cup \{x_1\} & {\rm if} \ x_2 \in \Phi \\
\Phi & {\rm if} \ x_2 \not \in \Phi
\end{array}
\right .
\\
\end{array}
}
{\Psi, \Phi, \Lambda \vdash x_1 := x_2: \Phi', \Lambda'}
\\ \ \\
(3) & \irule{
\Lambda' = \Lambda[x \mapsto (\interval{0}{\infty}, \automaton^*)]
}
{\Psi, \Phi, \Lambda \vdash x := \ ?: \Phi \setminus \{x\}, \Lambda'}  \\ \ \\
(4) & \irule{
\begin{array}{c}
\automaton = \emph{snd}(\Lambda(x)) \cap \evalregex{\impure}{\Lambda}\\
\Lambda' = \Lambda[x \mapsto (\emph{fst}(\Lambda(x)) , \automaton)]
\end{array}
}
{\Psi, \Phi, \Lambda \vdash {\rm assume}(x \in \impure): \Phi, \Lambda'}  \\ \ \\

(5) & \irule{
\begin{array}{c}
\interval{l_1}{u_1} =\emph{fst}(\Lambda(x))\\
\interval{l_2}{u_2} = \evalint{\nu}{\Lambda}\\
\interv = \interval{l_1}{\emph{min}(u_1, u_2)} \\
\Lambda' = \Lambda[x \mapsto ( \interv, \emph{snd}(\Lambda(x)))]
\end{array}
}
{\Psi, \Phi, \Lambda \vdash {\rm assume}(\len{x} \leq \nu): \Phi, \Lambda'}  \\ \ \\

(6) & \irule{
\begin{array}{c}
\automaton = \emph{snd}(\Psi(\regex)) \cap \emph{snd}({\Lambda(x)}) \\
x \not \in \Phi \lor \automaton = \automaton^\emptyset \lor \emph{fst}(\Psi(\regex)) \not \in \emph{fst}({\Lambda(x)})
\end{array}
}
{\Psi, \Phi, \Lambda \vdash {\rm match}(x, \regex): \Phi, \Lambda}  \\ \ \\
(7) & \irule{
\begin{array}{c}
\Psi, \Phi, \Lambda \vdash S_1: \Phi_1, \Lambda_1 \\
\Psi, \Phi_1, \Lambda_1 \vdash S_2: \Phi_2, \Lambda_2
\end{array}
}{\Psi, \Phi, \Lambda \vdash S_1; S_2: \Phi_2, \Lambda_2} \\ \ \\
(8) & \irule{
\begin{array}{c}
\Psi, \Phi, \Lambda \vdash S_1: \Phi_1, \Lambda_1 \\
\Psi, \Phi, \Lambda \vdash S_2: \Phi_2, \Lambda_2 \\
\Phi' = \Phi_1 \cup \Phi_2, \ \Lambda'= \Lambda_1 \sqcup \Lambda_2
\end{array}
}{\Psi, \Phi, \Lambda \vdash {\rm if}(\star) \ {\rm then} \ S_1 \ {\rm else} \ S_2:  \Phi', \Lambda' } \\ \ \\
(9) &
\irule{
\begin{array}{c}
\Phi^* \supseteq \Phi, \ \Lambda^* \sqsupseteq \Lambda \\
\Psi, \Phi^*, \Lambda^* \vdash S: \Phi^*, \Lambda^*
\end{array}
}
{\Psi, \Phi, \Lambda \vdash {\rm while}(\star) \ {\rm do} \ S: \Phi^*, \Lambda^*}
\end{array}
\]
\caption{Inference rules describing static analysis}
\label{fig:rules}
\end{figure}

In this figure, rule (1) describes the analysis of sources (i.e., $\emph{getInput}(x))$. Since variable $x$ is now tainted, we add it to our taint abstraction $\Phi$. Furthermore, since the user is free to supply any string,  $\Lambda'$ abstracts the length of $x$ using the interval $\interval{0}{\infty}$ and its contents using the automaton ~$\automaton^*$, which accepts any string.

Rule (2) for processing assignments $x_1 := x_2$ is straightforward: In particular, $x_1$ becomes tainted iff $x_2$ is tainted, and the string abstraction of $x_1$ is the same as variable $x_2$. For assignments of the form $x := ?$ (rule 3), we untaint variable $x$ by removing it from $\Phi$ since $?$ denotes string constants in \il. However, since $?$ represents unknown strings, $\Lambda'$ maps $x$ to $(\interval{0}{\infty}, \automaton^*)$.

Rule (4) describes the analysis of assumptions of the form $\emph{assume}(x \in \impure)$. Because $\impure$ can refer to program variables, we must first figure out the regular expressions that are represented by $\impure$. For this purpose, Fig.~\ref{fig:eval-impure} describes the evaluation of  impure regular expression $\impure$ under string abstraction $\Lambda$, denoted as $\evalregex{\impure}{\Lambda}$. Since the assumption states that the value stored in $x$ must be in the language $\evalregex{\impure}{\Lambda}$, the new string abstraction $\Lambda'$ maps $x$ to the automaton $\emph{snd}(\Lambda(x)) \cap \evalregex{\impure}{\Lambda}$.~\footnote{Observe that Rule (4) does not modify the length abstraction component of $\Lambda$. This is clearly sound, but potentially imprecise. However, since we model Java string operations by adding a pair of assumptions, one concerning length and the other concerning content, our analysis does not lead to a loss of precision because of the way assumptions are introduced.}

Rule (5), which is quite similar to rule (4), allows us to handle assumptions of the form $\emph{assume}(\emph{len}(x) \leq \nu)$. Since integer expression $\nu$ can refer to terms of the form $\emph{len}(y)$, we must evaluate $\nu$ under string abstraction $\Lambda$. For this purpose, Fig.~\ref{fig:eval-int} shows the evaluation of $\nu$ under $\Lambda$, denoted as $\evalint{\nu}{\Lambda}$. Now, going back to rule (5) of Fig.~\ref{fig:rules}, suppose that $\evalint{\nu}{\Lambda}$ yields the interval $\interval{l_2}{u_2}$, and suppose that $\Lambda$ maps $x$ to the length abstraction $\interval{l_1}{u_1}$. Clearly, the assumption $\emph{assume}(\emph{len}(x) \leq \nu)$ does not change the lower bound on $\emph{len}(x)$; hence the lower bound remains $l_1$. However, if $u_2$ is less than the previous upper bound $u_1$, we now have a more precise upper-bound $u_1$. Hence, the new string abstraction $\Lambda'$ maps the length component of $x$ to the interval $\interval{l_1}{\emph{min}(u_1, u_2)}$.

Rule (6) for $\emph{match}$ statements allows us to detect if the program contains a vulnerability. In particular, 
the premise of this rule states that either (1) $x$ is \emph{not} tainted ($x \not \in \Phi$ ) or (2) the automaton representing $x$'s contents does not contain any string in the attack automaton for $\regex$ (i.e., $\automaton = \automaton^\emptyset$), or (3) the length of $x$ cannot exceed the minimum bound given by $\Psi$ (i.e., $\emph{fst}(\Psi(\regex)) \not \in \emph{fst}(\Lambda(x))$). If these conditions in the premise of the \emph{match} rule are not met, then the program may contain a vulnerability.

The next rules  for sequencing (7) and conditionals (8) are fairly standard. Since we take the union of the taint abstractions in rule (7),  a variable $x$ becomes tainted if it was tainted in either branch of the conditional. Also, note that the join operator on string abstractions is defined in Fig.~\ref{fig:interval-ops}.
Finally, the last rule describes the analysis of loops. In particular,  rule (9) states that the abstractions $\Phi^*$ and $\Lambda^*$ overapproximate the behavior of the loop because (a) they subsume the initial abstractions $\Phi, \Lambda$ (first premise), and (b) they are inductive (second premise). 
While this rule does not describe how to compute $\Phi^*$ and $\Lambda^*$ in an algorithmic way, our implementation performs standard fixed point computation (using widening) to find these loop invariants.
 
 \begin{table*}[t]
\begin{center}
\scriptsize
\begin{tabular}{ |l|l| }
\hline
\scriptsize {\bf Java statement or predicate} & \hspace{1in} {\bf \il translation}  \\ \hline
{\tt x.contains(s)}  & 
$\emph{assume}(x \in (\Sigma^* \cdot s \cdot \Sigma^*)); \ \emph{assume}(\emph{len}(s) \leq \emph{len}(x))$ \\
\hline
{\tt y = x.replaceAll(a, b)}  & 
$(\ite{y:=x}{y:=?});  \  \emph{assume}(y \in (!a)^* ); \  \emph{assume}(\emph{len}(y) \leq \emph{len}(x))$ \\
\hline
{\tt y = x.substring(c1, c2)}  & 
$(\ite{y:=x}{y:=?});  \  \emph{assume}(\emph{len}(y) \leq c_2-c_1)$ \\
\hline
{\tt x.length() <= c}  & 
$\emph{assume}(\emph{len}(x) \leq c)$ \\
\hline
{\tt x.split(a).length() == c}  & 
$\emph{assume}(x \in ( (!a)^* \cdot a \cdot (!a)^*)^c )$ \\
\hline
{\tt x.indexOf(s) !=  -1}  & 
$\emph{assume}(x \in (\Sigma^* \cdot s \cdot \Sigma^*)); \ \emph{assume}(\emph{len}(s) \leq \emph{len}(x))$ \\
\hline
{\tt x.endsWith(y)}  & 
$\emph{assume}(x \in (\Sigma^* \cdot y)); \ \emph{assume}(\emph{len}(y) \leq \emph{len}(x))$ \\
\hline
{\tt x.equals(y)}  & 
$\emph{assume}(x \in y); \emph{assume}(\emph{len}(x) \leq \emph{len}(y));  \emph{assume}(\emph{len}(y) \leq \emph{len}(x))$ \\
\hline
{\tt x.matches($\regex$)}  & 
$\emph{assume}(x \in \regex);$ \\
\hline
{\tt x.startsWith(y)}  & 
$\emph{assume}(x \in (y \cdot \Sigma^*)); \ \emph{assume}(\emph{len}(y) \leq \emph{len}(x))$ \\
\hline
\end{tabular}
\end{center}
\vspace{-0.1in}
\caption{Examples illustrating translation from Java string operations to \il
  constructs. Here $x,y$ denote variables, $s$ denotes string constants,  $a, b$ represent
  distinct characters, and $c, c_1, c_2$ represent integer constants. The notation $!a$ means any character other than $a$, and $\ite{S_1}{S_2}$ is syntactic sugar for $\emph{if}(\star) \ \emph{then} \ S_1 \ \emph{else} \ S_2$. Observe that the statement $\ite{y:=x}{y:=?}$ has the effect of tainting $y$ if $x$ is tainted but does not introduce any assumptions about the content or size of string $y$.} \label{fig:translation}
\vspace{-0.1in}
\end{table*}

\fi\newpage

\end{document}